\documentclass[fleqn,usenatbib]{mnras}

\usepackage{newtxtext,newtxmath}
\usepackage[T1]{fontenc}

\DeclareRobustCommand{\VAN}[3]{#2}
\let\VANthebibliography\thebibliography
\def\thebibliography{\DeclareRobustCommand{\VAN}[3]{##3}\VANthebibliography}

\usepackage{graphicx}	% Including figure files
\usepackage{amsmath}	% Advanced maths commands
\usepackage{xspace}
\usepackage[export]{adjustbox}
\usepackage{natbib}
\usepackage{xspace}
\usepackage[table,xcdraw]{xcolor}
\usepackage{hyperref}
\usepackage{threeparttable}

\newcommand{\msun}{$M_\odot$\xspace}
\newcommand{\cratio}{$^{12}$C/$^{13}$C\xspace}

\title[\cratio Ratios in Open Cluster Red Giants]{An Investigation of Non-Canonical Mixing in Red Giant Stars Using APOGEE \cratio Ratios Observed in Open Cluster Stars}

\author[McCormick et al.]{
Caroline McCormick,$^{1}$\thanks{E-mail: uea6uk@virginia.edu} %orcid: 0000-0002-0963-2448
Steven R. Majewski,$^{1}$ %orcid: 0000-0003-2025-314
Verne V. Smith,$^{2,3}$ %orcid: 0000-0002-0134-2024
Christian R. Hayes,$^{4}$ %orcid: 0000-0003-2969-2445
Katia Cunha,$^{3,5,6}$ %orcid: 0000-0001-6476-0576
\newauthor Thomas Masseron,$^{7,8}$ %orcid: 0000-0002-6939-0831
Achim Weiss,$^{9}$ %orcid: 0000-0002-3843-1653
Matthew Shetrone,$^{10}$ %orcid: 0000-0003-0509-2656
Andr\'es Almeida,$^{1}$
Peter M. Frinchaboy,$^{11}$ %orcid: 0000-0002-0740-8346
\newauthor Domingo An\'ibal Garc\'ia-Hern\'andez,$^{7}$ %orcid: 0000-0002-1693-2721
and Christian Nitschelm$^{12}$ %orcid: 0000-0003-4752-4365
\\
$^{1}$Department of Astronomy, University of Virginia, 530 McCormick Road, Charlottesville, VA 22904, USA\\
$^{2}$NSF's National Optical-Infrared Astronomy Research Laboratory, 950 North Cherry Avenue, Tucson, AZ 85719, USA\\
$^{3}$Institut d'Astrophysique de Paris, UMR7095 CNRS, Sorbonne Universit\'e, 98bis Bd Arago, 75014 Paris, France\\
$^{4}$Herzberg Astronomy and Astrophysics Research Centre, 5071 West Saanich Road, Victoria, B.C., Canada, V9E 2E7\\
$^{5}$Observat\'orio Nacional, 77 Rua General Jos\'e Cristino, Rio de Janeiro, 20921-400, Brazil\\
$^{6}$Steward Observatory, University of Arizona, 933 North Cherry Avenue, Tucson, AZ 85721, USA\\
$^{7}$Instituto de Astrof\'isica de Canarias, 38205 La Laguna, Tenerife, Spain\\
$^{8}$Departamento de Astrof\'isica, Universidad de La Laguna, E-38206 La Laguna, Tenerife, Spain\\
$^{9}$Max-Planck-Institut f\"ur Astrophysik, Karl-Schwarzschild-Str. 1, D-85748 Garching, Germany\\
$^{10}$University of California Observatories, University of California Santa Cruz, Santa Cruz, CA 95064, USA\\
$^{11}$Department of Physics \& Astronomy, Texas Christian University, Fort Worth, TX 76129, USA\\
$^{12}$Centro de Astronom{\'i}a (CITEVA), Universidad de Antofagasta, Avenida Angamos 601, Antofagasta 1270300, Chile
}

\date{Accepted XXX. Received YYY; in original form ZZZ}

\pubyear{2022}

\begin{document}
\label{firstpage}
\pagerange{\pageref{firstpage}--\pageref{lastpage}}
\maketitle

% Abstract of the paper
\begin{abstract}
Standard stellar evolution theory poorly predicts the surface abundances of chemical species in low-mass, red giant branch (RGB) stars. Observations show an enhancement of p-p chain and CNO cycle products in red giant envelopes, which suggests the existence of non-canonical mixing that brings interior burning products to the surface of these stars. The \cratio ratio is a highly sensitive abundance metric used to probe this mixing. We investigate extra RGB mixing by examining (1) how \cratio is altered along the RGB and (2) how \cratio changes for stars of varying age and mass. Our sample consists of 43 red giants spread over 15 open clusters from the Sloan Digital Sky Survey's APOGEE DR17 that have reliable \cratio ratios derived from their APOGEE spectra. We vetted these \cratio ratios and compared them as a function of evolution and age/mass to the standard mixing model of stellar evolution and to a model that includes prescriptions for RGB thermohaline mixing and stellar rotation. We find that the observations deviate from standard mixing models, implying the need for extra mixing. Additionally, some of the abundance patterns depart from the thermohaline model, and it is unclear whether these differences are due to incomplete observations, issues inherent to the model, our assumption of the cause of extra mixing, or any combination of these factors. Nevertheless, the surface abundances across our age/mass range clearly deviate from the standard model, agreeing with the notion of a universal mechanism for RGB extra mixing in low-mass stars.
\end{abstract}


%%%%%%%%%%%%%%%%%%%%%%%%%%%%%%%%%%%%%%%%%%%%%%%%%%

%%%%%%%%%%%%%%%%% BODY OF PAPER %%%%%%%%%%%%%%%%%%

\section{Introduction}
\label{sec:intro}

\noindent A thorough knowledge of the chemical evolution of stellar populations, galaxies, and the universe as a whole is only achievable with a complete, or at least sound, understanding of stellar evolution through all major developmental phases for all initial stellar masses.  Particularly, in regards to the chemical evolution of galaxies and their interstellar media (ISM), it is essential to understand how the observed elemental abundance patterns in stars relate to their internal nucleosynthetic processes and the eventual yields they contribute to the ISM through chemical enrichment. 

One specific area of uncertainty is the array of mixing processes that take place in the interiors of evolved, low- and intermediate-mass stars\footnote{Here, low-mass stars are $\backsim$0.8--2 \msun stars, and intermediate-mass stars are $\backsim$2--8 \msun.} and how these processes affect surface abundances. Traditional stellar evolution models (i.e., models where only convection is responsible for interior mixing) predict that the surface abundances in low- and intermediate-mass, evolved stars should remain unchanged after the first dredge-up at the beginning of the red giant branch (RGB) until subsequent alterations take place during the asymptotic giant branch (AGB) phase. However, observations of upper RGB and horizontal branch stars (e.g., \citealt{SNEDEN86}; \citealt{GILROY89}; \citealt{GRATTON00}; \citealt{SMILJANIC09}; \citealt{TAUTVAISIENE10, TAUTVAISIENE13}; \citealt{DRAZDAUSKAS16}; \citealt{TAKEDA19}; \citealt{CHARBONNEL20}) have shown that certain surface abundances are in fact altered during this period of stellar evolution, which suggests the existence of a non-canonical mixing process at work.

A variety of physical mechanisms, such as cool bottom processing (\citealt{BSW95}; \citealt{WBS95}; \citealt{BS99}; \citealt{SB99}), stellar rotation (\citealt{SWEIGART79}; \citealt{CHARBONNEL95}; \citealt{CHANAME05}; \citealt{PALACIOS06}), and magnetic fields (\citealt{BUSSO07}; \citealt{DENISSENKOV09}), have been proposed to have at least some level of contribution to this non-canonical mixing on the upper RGB. However, there is still no definitive consensus on the exact conditions and processes at work that cause the extra mixing. One of the more popular physical mechanisms to which extra mixing is attributed (and one of the mechanisms to which we compare our data) is a double-diffusive instability generically referred to as a thermohaline instability in the literature (\citealt{STERN60}). \citet{CZ07} identified that this double-diffusive instability is the first instability to occur and alter the interior mean molecular weight ($\mu$) profile due to the growing inverse-$\mu$ gradient at this phase of evolution. Furthermore, this instability occurs naturally in low-mass and less massive intermediate-mass ($\lesssim$ 2.2 \msun; \citealt{CL10}) stars on the RGB. Stellar rotation is the second mechanism to which we compare our data, and it is also known to complicate the surface abundances in RGB stars. This stellar rotation during the main sequence causes the diffusion of material within a star, thereby changing the internal abundance profiles of species such as $^{12}$C and $^{13}$C. While these composition changes are not significant enough to produce noticeable changes at the surface of the star during the main sequence, the effects do show up during the first dredge-up when the envelope makes contact with the mixed interior regions (e.g., \citealt{PALACIOS03}; \citealt{CL10}).

During the first dredge-up, the convective envelope of the star reaches deep into regions that have been chemically modified by hydrogen burning and mixes to the surface matter enriched in by-products of the p-p chains and CNO cycle, such as $^3$He, $^{13}$C, $^{14}$N, and depleted in $^{12}$C and $^{7}$Li, thereby diluting the initial surface abundances of the star. The first dredge-up homogenizes the chemical composition of the red giant envelope and leaves behind a chemical discontinuity at the border between the farthest inward extent of the envelope during the first dredge-up and the radiative layer just outside the hydrogen-burning shell (HBS). Further along the RGB, the star reaches the so-called ``luminosity bump'' where the outward-extending HBS reaches the chemical discontinuity and causes a temporary dip in the stellar luminosity. It is at this point that the proposed thermohaline instability sets in to eventuate an extra mixing episode that further alters the surface abundances and produces some of the unusual patterns that have been observed.

Thermohaline instability is initiated by the $^3$He($^3$He, 2p)$^4$He p-p chain reaction occurring in the outer HBS. This reaction decreases $\mu$ in the burning region since more particles result from this reaction than the number of particles that were present initially. Therefore, $\mu$ increases outward, producing an inverse $\mu$ gradient locally. The higher-$\mu$ material sinks, and it is eventually mixed with its surroundings. As a result of this process, products of CNO burning such as $^{13}$C and $^{14}$N located in surrounding regions are transported throughout the thermohaline unstable region. Provided there is enough $^3$He to sustain this inverse $\mu$ gradient, the thermohaline unstable region will eventually come into contact with the convective envelope, causing further mixing of the burning products to the surface.

Because extra mixing on the RGB is directly connected to changes in surface abundances of p-p and CNO species, one way to probe the effects of the mixing is to compare the abundances and ratios of certain atomic species, such as \cratio or [C/N], between otherwise similar stars that are in evolutionary stages before, during, and after this mixing episode is expected to occur (e.g., \citealt{SZIGETI18}). Comparing these observations to models including prescriptions for the physical mechanism(s) (e.g., thermohaline instability, rotation) that could cause the extra mixing will help us better understand the interior mixing conditions in these stars. The present analysis relies on the \cratio ratio because it shows a heightened sensitivity to mixing. The ratio typically drops from $\gtrsim$70 to $\backsim$20 during the first dredge-up and to $\backsim$10 after extra mixing. Also, when compared to [C/N], the \cratio ratio is thought to be a more powerful tool to use in constraining extra mixing (see \citealt{LAGARDE19}). 

In this work, we employ data from Data Release 17 (DR17; \citealt{ABDURROUF22}) of the Sloan Digital Sky Survey IV's (SDSS-IV; \citealt{BLANTON17}) Apache Point Observatory Galactic Evolution Experiment (APOGEE; \citealt{MAJEWSKI17}) and its value added catalogs (VACs), which contain open cluster membership evaluations for $\backsim$26,000 stars and derived \cratio ratios for $\backsim$120,000 red giants. With this data, we obtain a sample of 43 confirmed open cluster red giant members with homogeneously derived \cratio ratios. Adopting open cluster stars for our analyses allows us to assign reliable ages\footnote{It is assumed that stars belonging to an open cluster all formed at the same time and are therefore the same age.} and initial masses for the stars belonging to each cluster. Our goal is to gain insight into the overall importance and cause of extra mixing as well as its effects in stars of varying age and mass on the RGB. Specifically, we study how \cratio changes over time and as a function of age and mass and compare these observations to models including the effects of thermohaline extra mixing and stellar rotation, for which models are publicly available for testing \citep{LAGARDE12}.

This paper is organized as follows: Section \ref{sec:data} outlines the APOGEE data used and the justification of the selection criteria applied to obtain our final sample of open cluster red giants. We present the evolution of the \cratio ratio along the RGB and the \cratio ratio as a function of age and mass in Section \ref{sec:results}. In Section \ref{sec:discussion}, we discuss the broader impact of our results with respect to stellar evolution and extra mixing model predictions, and finally, in Section \ref{sec:conclusions}, we summarize and draw conclusions from our work.

\section{Data}
\label{sec:data}

\noindent We utilize spectroscopic data from SDSS DR17 \citep{ABDURROUF22}---the final data release of SDSS-IV \citep{BLANTON17} collaboration. This data release contains all of the data taken as part of the APOGEE and APOGEE-2 surveys \citep{MAJEWSKI17} which used the two APOGEE spectrographs \citep{wilson2019}: APOGEE-N on the Sloan 2.5-meter Telescope in New Mexico \citep{Gunn2006} with an auxiliary feed from the NMSU 1-meter telescope (\citealt{Holtzman2010}) and APOGEE-S on the 2.5-meter du Pont Telescope \citep{bv73} at Las Campanas Observatory in Chile. Targeting for the APOGEE survey is described in \citet{zas13}, while that for the APOGEE-2 survey is described in \citet{zas17}, \citet{Beaton2021}, and \citet{Santana2021}. Additionally, \citet{FRINCHABOY13} and \citet{DONOR18} give targeting information for the open clusters observed in APOGEE. The data reduction pipeline for APOGEE is described in \citet{dln15} and in \citet{Holtzman2015} for the APOGEE spectra taken with the 1-meter telescope.

The APOGEE Stellar Atmospheric Parameters and Chemical Abundances Pipeline (ASPCAP; \citealt{GarciaPerez2016}), which is based on the FERRE code written by \citet{AllendePrieto2006}, obtains stellar atmospheric parameters and chemical abundances by finding the best match in a library of synthetic spectra. For DR17, ASPCAP uses a grid of MARCS stellar atmospheres \citep{GUSTAFSSON08,JONSSON20}, and an $H$-band line list from \citet{Smith2021}, which is an update of the \citet{shetrone2015} line list. 

\begin{figure}
\centering
\includegraphics[scale=0.5]{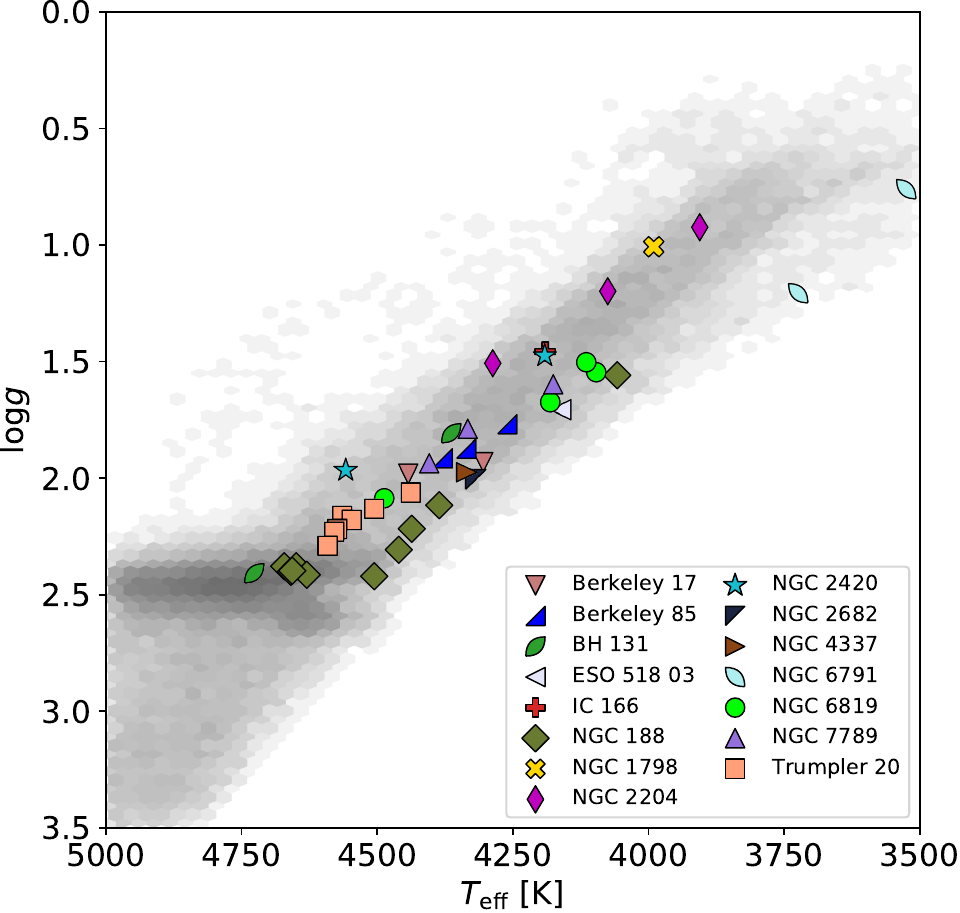}
\caption{The $T_{\rm eff}$ - $\log{g}$ diagram of all cluster stars in our final sample ({\it colored symbols}). The background gray scale shows the relative density of stars in the BAWLAS VAC with derived \cratio.}
\label{fig:logg_Teff_combined}
\end{figure}

{
\renewcommand{\arraystretch}{1.25}
\begin{table*}
%\begin{threeparttable}
\centering
\caption{Open clusters with at least one red giant with a reliable \cratio ratio derived from the BAWLAS VAC. Also listed are the cluster mean metallicities ([Fe/H]) and standard error, the initial stellar mass of stars at the cluster main sequence turn off, the adopted cluster age, and the literature source for the age of each cluster.}
\begin{tabular}{ccccccc}
\hline \hline
\multicolumn{1}{m{1.75cm}}{\centering Cluster Name} & \multicolumn{1}{m{1cm}}{\centering Number of Stars} & \multicolumn{1}{m{2.3cm}}{\centering Number of BAWLAS \cratio Limit Stars} & 
\multicolumn{1}{m{1.75cm}}{\centering{[}Fe/H{]}} &
\multicolumn{1}{m{1.75cm}}{\centering Mass \\ (\msun)} & \multicolumn{1}{m{1.75cm}}{\centering Age \\ (Gyr)} & \multicolumn{1}{m{4cm}}{\centering Age Source} \\ \hline
Berkeley 17 & 2 & 2 & -0.17 $\pm$ 0.01 & $1.05${\raisebox{0.5ex}{\tiny$^{+0.08}_{-0.11}$}} & $7.24${\raisebox{0.5ex}{\tiny$^{+2.99}_{-2.11}$}} & \citet{CG20} \\
Berkeley 85 & 4 & 1 & 0.10 $\pm$ 0.01 & $2.91${\raisebox{0.5ex}{\tiny$^{+0.43}_{-0.61}$}} & $0.42${\raisebox{0.5ex}{\tiny$^{+0.17}_{-0.12}$}} & \citet{CG20} \\
BH 131 & 2 & 3 & 0.09 $\pm$ 0.01 & 1.93 & 1.26 & \citet{DIAS02}$^{1}$\tnote{1} \\
ESO 518 03 & 1 & 2 & 0.03 $\pm$ 0.01 & $1.83${\raisebox{0.5ex}{\tiny$^{+0.21}_{-0.30}$}} & $1.41${\raisebox{0.5ex}{\tiny$^{+0.59}_{-0.41}$}} & \citet{CG20} \\
IC 166 & 1 & 0 & -0.16 $\pm$ 0.01 & $1.79${\raisebox{0.5ex}{\tiny$^{+0.22}_{-0.30}$}} & $1.32${\raisebox{0.5ex}{\tiny$^{+0.54}_{-0.41}$}} & \citet{CG20} \\
NGC 188 & 10 & 6 & 0.06 $\pm$ 0.01 & $1.11${\raisebox{0.5ex}{\tiny$^{+0.08}_{-0.11}$}} & $7.08${\raisebox{0.5ex}{\tiny$^{+2.92}_{-2.07}$}} & \citet{CG20} \\
NGC 1798 & 1 & 3 & -0.35 $\pm$ 0.01 & $1.58${\raisebox{0.5ex}{\tiny$^{+0.19}_{-0.26}$}} & $1.66${\raisebox{0.5ex}{\tiny$^{+0.68}_{-0.49}$}} & \citet{CG20} \\
NGC 2204 & 3 & 6 & -0.36 $\pm$ 0.01 & $1.45${\raisebox{0.5ex}{\tiny$^{+0.16}_{-0.23}$}} & $2.09${\raisebox{0.5ex}{\tiny$^{+0.86}_{-0.61}$}} & \citet{CG20} \\
NGC 2420 & 2 & 5 & -0.26 $\pm$ 0.01 & $1.58${\raisebox{0.5ex}{\tiny$^{+0.18}_{-0.26}$}} & $1.74${\raisebox{0.5ex}{\tiny$^{+0.71}_{-0.51}$}} & \citet{CG20} \\
NGC 2682 & 1 & 18 & -0.03 $\pm$ 0.01 & $1.25${\raisebox{0.5ex}{\tiny$^{+0.11}_{-0.15}$}} & $4.27${\raisebox{0.5ex}{\tiny$^{+1.76}_{-1.25}$}} & \citet{CG20} \\
NGC 4337 & 1 & 3 & 0.19 $\pm$ 0.01 & $1.88${\raisebox{0.5ex}{\tiny$^{+0.21}_{-0.28}$}} & $1.45${\raisebox{0.5ex}{\tiny$^{+0.59}_{-0.43}$}} & \citet{CG20} \\
NGC 6791 & 4 & 0 & 0.28 $\pm$ 0.02 & $1.10${\raisebox{0.5ex}{\tiny$^{+0.001}_{-0.001}$}} & $8.45${\raisebox{0.5ex}{\tiny$^{+0.04}_{-0.04}$}} & \citet{BOSSINI19} \\
NGC 6819 & 4 & 13 & -0.02 $\pm$ 0.02 & $1.53${\raisebox{0.5ex}{\tiny$^{+0.16}_{-0.22}$}} & $2.24${\raisebox{0.5ex}{\tiny$^{+0.92}_{-0.66}$}} & \citet{CG20} \\
NGC 7789 & 4 & 24 & -0.05 $\pm$ 0.01 & $1.73${\raisebox{0.5ex}{\tiny$^{+0.19}_{-0.28}$}} & $1.55${\raisebox{0.5ex}{\tiny$^{+0.64}_{-0.45}$}} & \citet{CG20} \\
Trumpler 20 & 9 & 2 & 0.08 $\pm$ 0.01 & $1.68${\raisebox{0.5ex}{\tiny$^{+0.17}_{-0.25}$}} & $1.86${\raisebox{0.5ex}{\tiny$^{+0.77}_{-0.54}$}} & \citet{CG20} \\ \hline
\end{tabular}
\begin{tablenotes}
\item [1] $^{1}$ No cluster age uncertainty was provided.
\end{tablenotes}
\label{tab:clusters}
%\end{threeparttable}
\end{table*}
}

\begin{figure*}
\centering
\includegraphics[scale=1.2]{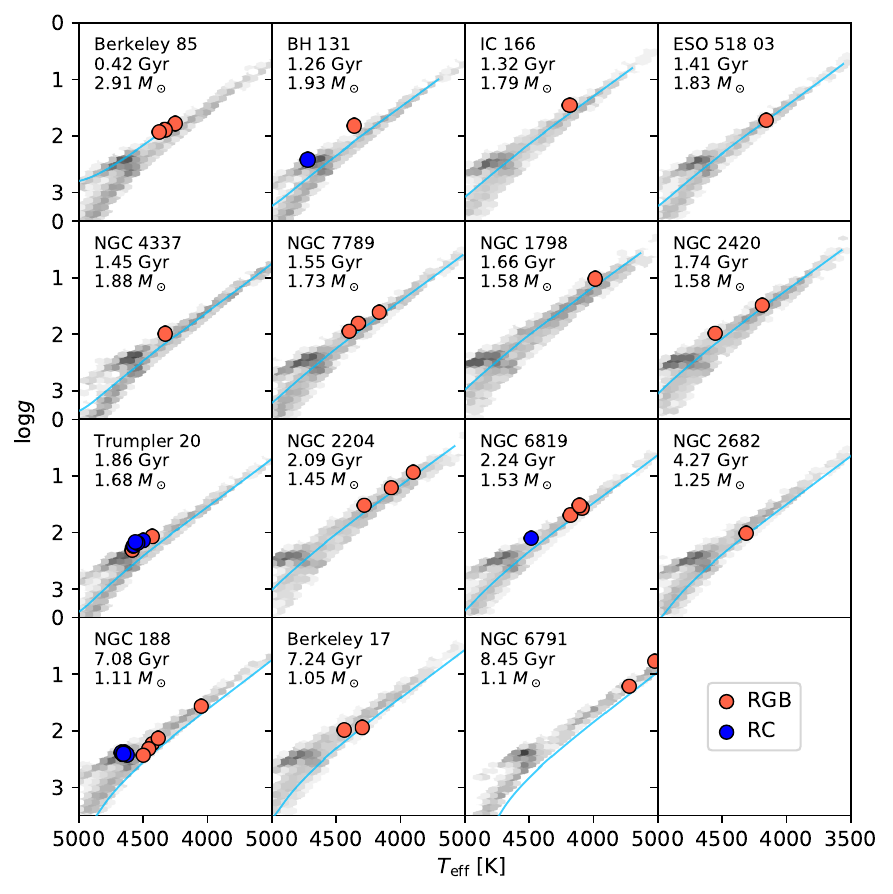}
\caption{The $T_{\rm eff}$ - $\log{g}$ diagram of our final sample stars ({\it circles}) in each cluster along with the best-matching cluster isochrone ({\it light blue curves}) generated using the MESA Isochrones \& Stellar Tracks (MIST) models (\citealt{DOTTER16}; \citealt{CHOI16}; \citealt{PAXTON11, PAXTON13, PAXTON15, PAXTON18}). Typical errors in $\log{g}$ are 0.02 dex and in $T_{\rm eff}$ are 3-8 K. {\it Blue circles} correspond to red clump stars that have ignited core helium burning, while {\it orange circles} correspond to stars on the RGB. The background gray scale shows the density of stars in the BAWLAS VAC with derived \cratio and [Fe/H] within 0.03 dex of the cluster mean [Fe/H].}
\label{fig:logg_Teff_grid}
\end{figure*}

For the present work, the \cratio ratios were derived from APOGEE DR17 spectra and stellar parameters from ASPCAP using the Brussels Automatic Code for Characterizing High accUracy Spectra (BACCHUS; \citealt{BACCHUS}); these ratios are reported in the BACCHUS Analysis of Weak-Lines in APOGEE Spectra (BAWLAS) VAC which contains data for $\backsim$120,000 red giants \citep{HAYES}. The stars analyzed in the BAWLAS VAC, including our final sample cluster stars, can be seen in the $T_{\rm eff}$ - $\log{g}$ diagrams in Figures \ref{fig:logg_Teff_combined} and \ref{fig:logg_Teff_grid}. We next describe the cuts and requirements applied to the full APOGEE data set to derive our final sample of red giants and their stellar parameters.

\subsection{Cluster Membership Cuts}
\label{subsec:membership}

\noindent We first sought red giants that are members of Galactic open clusters because ages and initial masses for these clusters and stars can be reliably inferred. To verify cluster membership, we used the Open Cluster Chemical Abundance and Mapping (OCCAM) survey (\citealt{DONOR18}; \citealt{DONOR20}; \citealt{MYERS}), which provides cluster membership probabilities for 26,699 stars in 153 open clusters observed in APOGEE. Of the 153 open clusters, we only considered the best clusters as denoted by the quality flag given in OCCAM being set to 1 or 2 (see \citealt{DONOR20} for definition). Additionally, we required that each cluster have at least five reliably determined member stars identified in OCCAM to provide a greater chance at having well-populated clusters in our analyses and to have a higher confidence in the membership analysis for each cluster.

We then analyzed the membership probability for each star supposedly belonging to each of these clusters. To be a cluster member according to the OCCAM survey, a star must have a radial velocity (RV), metallicity ([Fe/H]), and proper motion (PM) within three standard deviations of the cluster mean values. In other words, the ``RV Prob,'' ``[Fe/H] Prob,'' and ``PM Prob'' reported values must be $>$0.01.

From the combination of these open cluster membership criteria, the initial OCCAM sample of 26,699 stars is reduced to 1,196 reliable cluster members belonging to 43 clusters.

\subsection{BAWLAS VAC Carbon Measurement Criteria}
\label{subsec:c12c13_cut}

\noindent Due to the difficulty of measuring the weak lines that are used in determining the \cratio ratio, the BAWLAS VAC includes trustworthy \cratio ratios for 52,855 of its stars and \cratio lower, or $^{13}$C upper, limits for 49,252 stars. For a star to be included in our final sample, we required that the star must have a non-limit \cratio value. Combining this criterion with the verified open cluster member stars from the OCCAM survey, there are 212 stars belonging to 24 open clusters.

\subsection{Age and Mass Determinations}

\noindent Because we are investigating the \cratio ratio as a function of age, we further limited the sample of stars to only include stars in clusters with previously determined ages. We surveyed the literature for open cluster ages, seeking to find a set of estimates where most, if not all, cluster ages in our sample are determined in a consistent manner. No single source was found that had reported ages for all of the clusters arising from the \S\ref{subsec:membership} OCCAM membership and \S\ref{subsec:c12c13_cut} \cratio cuts. However, we minimized the variety of sources by adopting cluster ages from three sources: \citet{CG20}, \citet{BOSSINI19}, and \citet{DIAS02}. \citet{CG20} was our default cluster age source, as they provide consistently derived and generally reliable cluster ages for a large number of clusters. We used \citet{BOSSINI19} and \citet{DIAS02} when ages from \citet{CG20} were either untrustworthy (NGC 6791; \citealt{BROGAARD21}) or unavailable (BH 131). We found that 15 of the 24 open clusters have ages reported by these sources, resulting in a sample of 49 red giant stars in these 15 particular clusters. 

Table \ref{tab:clusters} lists the final collection of clusters represented in our sample, along with the number of stars in each cluster with reliable \cratio ratios and \cratio limits, the DR17 mean cluster metallicity ([Fe/H]) derived from ASPCAP, the age of each cluster, the literature source for each age, and the initial stellar mass for RGB stars in each cluster. These masses were determined from MESA Isochrones \& Stellar Tracks (MIST) isochrones (\citealt{DOTTER16}; \citealt{CHOI16}; \citealt{PAXTON11, PAXTON13, PAXTON15, PAXTON18}) which adopt solar-scaled abundances. We input the cluster's age and mean [Fe/H] and adopted the initial mass of a star at the terminal age main sequence, which is at equivalent evolutionary point (EEP) 454, as the red giant initial mass for each cluster. We acquired masses for RGB stars in all 15 clusters with reported ages, so no further cuts are made to the sample here.

\subsection{Spectral Fit Cuts}

\begin{table*}
\centering
\caption{Stellar parameters and manually vetted \cratio ratios for all open cluster red giants (43) in the BAWLAS VAC that meet the selection criteria described in \S\ref{sec:data}. Also listed are the radial velocity, [Fe/H], and proper motion probabilities from OCCAM used to determine cluster membership for each star and the evolutionary state for each star as reported in APOGEE.}
\label{tab:stars}
\begin{tabular}{cccccccccc}
\hline \hline
\multicolumn{1}{m{2.5cm}}{\centering APOGEE ID} & \multicolumn{1}{m{1.5cm}}{\centering Cluster} & \multicolumn{1}{m{1.5cm}}{\centering T$_{\rm eff}$ \\ (K)} & \multicolumn{1}{m{1.5cm}}{\centering $\log{g}$} & \multicolumn{1}{m{1.5cm}}{\centering [Fe/H]} & \multicolumn{1}{m{1cm}}{\centering \cratio} & \multicolumn{1}{m{0.75cm}}{\centering RV Prob} & \multicolumn{1}{m{0.75cm}}{\centering [Fe/H] Prob} & \multicolumn{1}{m{0.75cm}}{\centering PM Prob} & \multicolumn{1}{m{0.75cm}}{\centering Evol. State} \\ \hline
2M05203799+3034414 & Berkeley 17 & 4307 $\pm$ 6 & 1.93 $\pm$ 0.02 & -0.18 $\pm$ 0.01 & 13 $\pm$ 1 & 1.00 & 0.98 & 0.97 & RGB\\
2M05203650+3030351 & Berkeley 17 & 4445 $\pm$ 7 & 1.98 $\pm$ 0.02 & -0.17 $\pm$ 0.01 & 11 $\pm$ 2 & 0.82 & 0.90 & 0.94 & RGB\\
2M20183476+3740565 & Berkeley 85 & 4380 $\pm$ 6 & 1.91 $\pm$ 0.02 & 0.09 $\pm$ 0.01 & 16 $\pm$ 1 & 0.87 & 1.00 & 0.84 & RGB\\
2M20183785+3743009 & Berkeley 85 & 4337 $\pm$ 6 & 1.87 $\pm$ 0.02 & 0.12 $\pm$ 0.01 & 17 $\pm$ 4 & 0.89 & 0.72 & 0.96 & RGB\\
2M20184497+3744174 & Berkeley 85 & 4262 $\pm$ 5 & 1.77 $\pm$ 0.02 & 0.09 $\pm$ 0.01 & 14 $\pm$ 3 & 0.97 & 0.98 & 0.65 & RGB\\
2M12260433-6324196 & BH 131 & 4365 $\pm$ 6 & 1.81 $\pm$ 0.02 & 0.09 $\pm$ 0.01 & 15 $\pm$ 2 & 0.87 & 0.95 & 0.99 & RGB\\
2M12261653-6325258 & BH 131 & 4728 $\pm$ 8 & 2.41 $\pm$ 0.02 & 0.08 $\pm$ 0.01 & 12 $\pm$ 3 & 0.82 & 0.80 & 1.00 & RC\\
2M16464504-2558201 & ESO 518 03 & 4163 $\pm$ 5 & 1.71 $\pm$ 0.02 & 0.03 $\pm$ 0.01 & 15 $\pm$ 3 & 0.98 & 0.91 & 0.85 & RGB\\
2M01522919+6159381 & IC 166 & 4191 $\pm$ 6 & 1.45 $\pm$ 0.02 & -0.16 $\pm$ 0.01 & 14 $\pm$ 2 & 0.46 & 0.26 & 2.00 & RGB\\
2M00455119+8518082 & NGC 188 & 4461 $\pm$ 6 & 2.31 $\pm$ 0.02 & 0.05 $\pm$ 0.01 & 16 $\pm$ 4 & 0.35 & 0.91 & 0.26 & RGB\\
2M00441241+8509312 & NGC 188 & 4059 $\pm$ 5 & 1.56 $\pm$ 0.02 & -0.01 $\pm$ 0.01 & 17 $\pm$ 4 & 1.00 & 0.14 & 0.55 & RGB\\
2M00320079+8511465 & NGC 188 & 4507 $\pm$ 6 & 2.42 $\pm$ 0.02 & 0.07 $\pm$ 0.01 & 13 $\pm$ 2 & 0.98 & 0.99 & 0.05 & RGB\\
2M00465966+8513157 & NGC 188 & 4650 $\pm$ 7 & 2.38 $\pm$ 0.02 & 0.04 $\pm$ 0.01 & 17 $\pm$ 5 & 0.94 & 0.78 & 0.71 & RC\\
2M00350924+8517169 & NGC 188 & 4673 $\pm$ 7 & 2.38 $\pm$ 0.02 & 0.05 $\pm$ 0.01 & 12 $\pm$ 1 & 0.98 & 0.91 & 0.12 & RC\\
2M00571844+8510288 & NGC 188 & 4631 $\pm$ 7 & 2.41 $\pm$ 0.02 & 0.09 $\pm$ 0.01 & 13 $\pm$ 0.3 & 1.00 & 0.85 & 0.83 & RC\\
2M00415197+8527070 & NGC 188 & 4661 $\pm$ 7 & 2.40 $\pm$ 0.02 & 0.08 $\pm$ 0.01 & 12 $\pm$ 1 & 0.92 & 0.93 & 0.32 & RC\\
2M00445253+8514055 & NGC 188 & 4437 $\pm$ 6 & 2.22 $\pm$ 0.02 & 0.04 $\pm$ 0.01 & 10 $\pm$ 3 & 0.80 & 0.71 & 0.79 & RGB\\
2M00581691+8540183 & NGC 188 & 4658 $\pm$ 7 & 2.40 $\pm$ 0.02 & 0.11 $\pm$ 0.01 & 7 $\pm$ 2 & 0.88 & 0.80 & 0.75 & RC\\
2M00463920+8523336 & NGC 188 & 4387 $\pm$ 6 & 2.12 $\pm$ 0.02 & 0.04 $\pm$ 0.01 & 8 $\pm$ 2 & 0.98 & 0.74 & 0.03 & RGB\\
2M05114795+4740258 & NGC 1798 & 3991 $\pm$ 5 & 1.01 $\pm$ 0.02 & -0.35 $\pm$ 0.01 & 15 $\pm$ 3 & 1.00 & 0.04 & 0.51 & RGB\\
2M06153140-1842562 & NGC 2204 & 4077 $\pm$ 5 & 1.20 $\pm$ 0.02 & -0.35 $\pm$ 0.01 & 12 $\pm$ 1 & 0.88 & 0.41 & 0.78 & RGB\\
2M06145845-1838429 & NGC 2204 & 4289 $\pm$ 6 & 1.50 $\pm$ 0.02 & -0.34 $\pm$ 0.01 & 11 $\pm$ 2 & 1.00 & 0.45 & 0.72 & RGB\\
2M06153666-1846527 & NGC 2204 & 3907 $\pm$ 5 & 0.93 $\pm$ 0.02 & -0.38 $\pm$ 0.01 & 11 $\pm$ 1 & 0.98 & 0.07 & 0.94 & RGB\\
2M07381507+2134589 & NGC 2420 & 4194 $\pm$ 6 & 1.48 $\pm$ 0.02 & -0.27 $\pm$ 0.01 & 10 $\pm$ 0.6 & 1.00 & 0.16 & 0.99 & RGB\\
2M07382166+2133514 & NGC 2420 & 4559 $\pm$ 7 & 1.97 $\pm$ 0.02 & -0.25 $\pm$ 0.01 & 7 $\pm$ 0.1 & 0.99 & 0.34 & 0.02 & RGB\\
2M08493465+1151256 & NGC 2682 & 4320 $\pm$ 6 & 2.00 $\pm$ 0.02 & -0.03 $\pm$ 0.01 & 8 $\pm$ 1 & 0.99 & 0.71 & 0.73 & RGB\\
2M12240101-5807554 & NGC 4337 & 4336 $\pm$ 6 & 1.97 $\pm$ 0.02 & 0.19 $\pm$ 0.01 & 15 $\pm$ 4 & 1.00 & 0.81 & 0.78 & RGB\\
2M19213390+3750202 & NGC 6791 & 3724 $\pm$ 4 & 1.20 $\pm$ 0.02 & 0.29 $\pm$ 0.01 & 13 $\pm$ 3 & 0.99 & 0.77 & 2.00 & RGB\\
2M19211606+3746462 & NGC 6791 & 3527 $\pm$ 3 & 0.76 $\pm$ 0.02 & 0.23 $\pm$ 0.01 & 11 $\pm$ 1 & 0.98 & 0.10 & 0.61 & RGB\\
2M19411705+4010517 & NGC 6819 & 4098 $\pm$ 5 & 1.55 $\pm$ 0.02 & -0.03 $\pm$ 0.01 & 15 $\pm$ 2 & 0.73 & 0.26 & 0.93 & RGB\\
2M19411971+4023362 & NGC 6819 & 4116 $\pm$ 5 & 1.50 $\pm$ 0.02 & -0.06 $\pm$ 0.01 & 12 $\pm$ 2 & 0.82 & 0.05 & 2.00 & RGB\\
2M19413439+4017482 & NGC 6819 & 4183 $\pm$ 5 & 1.67 $\pm$ 0.02 & 0.02 $\pm$ 0.01 & 13 $\pm$ 1 & 1.00 & 0.91 & 0.40 & RGB\\
2M19412658+4011418 & NGC 6819 & 4488 $\pm$ 6 & 2.09 $\pm$ 0.02 & 0.00 $\pm$ 0.01 & 10 $\pm$ 2 & 0.98 & 0.65 & 0.28 & RC\\
2M23570744+5641417 & NGC 7789 & 4177 $\pm$ 5 & 1.60 $\pm$ 0.02 & -0.06 $\pm$ 0.01 & 17 $\pm$ 3 & 0.99 & 0.77 & 0.94 & RGB\\
2M23555312+5641203 & NGC 7789 & 4405 $\pm$ 6 & 1.93 $\pm$ 0.02 & -0.05 $\pm$ 0.01 & 11 $\pm$ 3 & 0.67 & 0.84 & 0.07 & RGB\\
2M23581471+5651466 & NGC 7789 & 4335 $\pm$ 6 & 1.79 $\pm$ 0.02 & -0.04 $\pm$ 0.01 & 10 $\pm$ 0.9 & 0.49 & 0.80 & 0.72 & RGB\\
2M12400451-6036566 & Trumpler 20 & 4440 $\pm$ 6 & 2.07 $\pm$ 0.02 & 0.06 $\pm$ 0.01 & 15 $\pm$ 2 & 0.66 & 0.28 & 0.90 & RGB\\
2M12390411-6034001 & Trumpler 20 & 4548 $\pm$ 7 & 2.18 $\pm$ 0.02 & 0.09 $\pm$ 0.01 & 10 $\pm$ 1 & 1.00 & 1.00 & 0.94 & RC\\
2M12400755-6035445 & Trumpler 20 & 4507 $\pm$ 6 & 2.13 $\pm$ 0.02 & 0.11 $\pm$ 0.01 & 10 $\pm$ 0.5 & 1.00 & 0.99 & 0.98 & RC\\
2M12400260-6039545 & Trumpler 20 & 4580 $\pm$ 7 & 2.23 $\pm$ 0.02 & 0.10 $\pm$ 0.01 & 10 $\pm$ 2 & 0.92 & 0.99 & 0.99 & RC\\
2M12391003-6038402 & Trumpler 20 & 4575 $\pm$ 7 & 2.22 $\pm$ 0.02 & 0.08 $\pm$ 0.01 & 11 $\pm$ 2 & 0.81 & 0.95 & 0.98 & RC\\
2M12402949-6038518 & Trumpler 20 & 4593 $\pm$ 7 & 2.29 $\pm$ 0.02 & 0.08 $\pm$ 0.01 & 11 $\pm$ 1 & 0.86 & 0.86 & 0.93 & RGB\\
2M12385807-6030286 & Trumpler 20 & 4566 $\pm$ 7 & 2.16 $\pm$ 0.02 & 0.07 $\pm$ 0.01 & 8 $\pm$ 2 & 0.81 & 0.48 & 0.65 & RC\\ \hline
\end{tabular}
\end{table*}

\begin{table*}
\centering
\caption{All stars (49) in our sample before vetting the \cratio values. Listed for comparison are the \cratio ratios determined in the BAWLAS VAC and those with manually vetted spectral fits of each star's lines. The ``Vetted Lines'' column shows the CN (i.e., 15641.7 \AA\ and partially 16121.4 \AA) and CO (i.e., partially 16121.4 \AA\ and the remaining lines) lines we used in our calculation of \cratio. The stars considered as \cratio limits in our vetting process are denoted by $>$ in the ``Manual \cratio'' column.}
\label{tab:C12C13_initial_final}
\begin{tabular}{cccccc}
\hline \hline
\multicolumn{1}{m{2.75cm}}{\centering APOGEE ID} & \multicolumn{1}{m{1.75cm}}{\centering Cluster} & \multicolumn{1}{m{8cm}}{\centering Vetted Lines \\ (\AA)} & 
\multicolumn{1}{m{1.5cm}}{\centering BAWLAS \cratio} & \multicolumn{1}{m{1.25cm}}{\centering Manual \\ \cratio}\\ \hline
2M05203799+3034414 & Berkeley 17 & 15641.7, 16121.4, 16530.0 & 13 $\pm$ 2 & 13 $\pm$ 1 \\
2M05203650+3030351 & Berkeley 17 & 15641.7, 16121.4, 16323.4, 16326.0, 16530.0, 16744.7 & 12 $\pm$ 4 & 11 $\pm$ 2 \\
2M20183476+3740565 & Berkeley 85 & 15641.7, 16121.4, 16530.0 & 20 $\pm$ 5 & 16 $\pm$ 1 \\
2M20183785+3743009 & Berkeley 85 & 15641.7, 16121.4, 16323.4, 16530.0 & 17 $\pm$ 6 & 17 $\pm$ 4 \\
2M20190397+3745002 & Berkeley 85 & 16121.4, 16530.0 & 14 $\pm$ 2 & $>$15 $\pm$ 2 \\
2M20184497+3744174 & Berkeley 85 & 15641.7, 16121.4, 16323.4, 16530.0 & 14 $\pm$ 3 & 14 $\pm$ 3 \\
2M12260433-6324196 & BH 131 & 15641.7, 16121.4, 16530.0 & 15 $\pm$ 2 & 15 $\pm$ 2 \\
2M12261653-6325258 & BH 131 & 15641.7, 16121.4, 16323.4, 16530.0, 16744.7 & 11 $\pm$ 2 & 12 $\pm$ 3 \\
2M16464504-2558201 & ESO 518 03 & 15641.7, 16326.0, 16530.0, 16744.7 & 15 $\pm$ 4 & 15 $\pm$ 3 \\
2M01522919+6159381 & IC 166 & 15641.7, 16121.4, 16530.0, 16744.7 & 12 $\pm$ 1 & 14 $\pm$ 2 \\
2M00455119+8518082 & NGC 188 & 15641.7, 16121.4, 16323.4, 16326.0, 16530.0 & 22 $\pm$ 5 & 16 $\pm$ 4 \\
2M00441241+8509312 & NGC 188 & 15641.7, 16323.4, 16326.0, 16530.0, 16741.2, 16744.7 & 15 $\pm$ 1 & 17 $\pm$ 4 \\
2M00320079+8511465 & NGC 188 & 15641.7, 16121.4, 16326.0, 16530.0 & 12 $\pm$ 2 & 13 $\pm$ 2 \\
2M00465966+8513157 & NGC 188 & 15641.7, 16530.0 & 12 $\pm$ 5 & 17 $\pm$ 5 \\
2M00350924+8517169 & NGC 188 & 15641.7, 16121.4, 16530.0, 16744.7 & 12 $\pm$ 3 & 12 $\pm$ 1 \\
2M00571844+8510288 & NGC 188 & 15641.7, 16121.4, 16530.0 & 12 $\pm$ 3 & 13 $\pm$ 0.3 \\
2M00415197+8527070 & NGC 188 & 15641.7, 16121.4, 16323.4, 16530.0 & 12 $\pm$ 4 & 12 $\pm$ 1 \\
2M00445253+8514055 & NGC 188 & 15641.7, 16121.4, 16323.4, 16326.0, 16530.0 & 9 $\pm$ 2 & 10 $\pm$ 3 \\
2M00581691+8540183 & NGC 188 & 15641.7, 16121.4, 16323.4, 16326.0, 16530.0 & 8 $\pm$ 1 & 7 $\pm$ 2 \\
2M00463920+8523336 & NGC 188 & 15641.7, 16121.4, 16326.0, 16530.0, 16744.7 & 6 $\pm$ 0.9 & 8 $\pm$ 2 \\
2M05114795+4740258 & NGC 1798 & 15641.7, 16121.4, 16323.4, 16326.0, 16530.0, 16741.2, 16744.7 & 14 $\pm$ 1 & 15 $\pm$ 3 \\
2M06153140-1842562 & NGC 2204 & 15641.7, 16121.4, 16326.0, 16530.0, 16744.7 & 13 $\pm$ 2 & 12 $\pm$ 1 \\
2M06145845-1838429 & NGC 2204 & 15641.7, 16121.4, 16323.4, 16326.0, 16327.3, 16530.0 & 12 $\pm$ 3 & 11 $\pm$ 2 \\
2M06153666-1846527 & NGC 2204 & 15641.7, 16121.4, 16323.4, 16326.0, 16530.0 & 11 $\pm$ 2 & 11 $\pm$ 1 \\
2M07381507+2134589 & NGC 2420 & 15641.7, 16326.0, 16530.0, 16744.7 & 10 $\pm$ 0.5 & 10 $\pm$ 0.6 \\
2M07382166+2133514 & NGC 2420 & 15641.7, 16530.0 & 7 $\pm$ 0.9 & 7 $\pm$ 0.1 \\
2M08493465+1151256 & NGC 2682 & 15641.7, 16121.4, 16326.0, 16530.0, 16744.7 & 8 $\pm$ 0.9 & 8 $\pm$ 1 \\
2M12240101-5807554 & NGC 4337 & 15641.7, 16323.4, 16530.0 & 13 $\pm$ 2 & 15 $\pm$ 4 \\
2M19212437+3735402 & NGC 6791 & 16326.0, 16744.7 & 24 $\pm$ 4 & $>$18 $\pm$ 2 \\
2M19213390+3750202 & NGC 6791 & 15641.7, 16327.3, 16530.0 & 12 $\pm$ 0.9 & 13 $\pm$ 3 \\
2M19211606+3746462 & NGC 6791 & 15641.7, 16121.4, 16530.0, 16741.2, 16744.7 & 11 $\pm$ 0.8 & 11 $\pm$ 1 \\
2M19213635+3739445 & NGC 6791 & 15641.7, 16121.4, 16744.7 & 10 $\pm$ 1 & $>$8 $\pm$ 1 \\
2M19411705+4010517 & NGC 6819 & 15641.7, 16121.4, 16530.0, 16744.7 & 18 $\pm$ 3 & 15 $\pm$ 2 \\
2M19411971+4023362 & NGC 6819 & 15641.7, 16121.4, 16530.0, 16744.7 & 13 $\pm$ 1 & 12 $\pm$ 2 \\
2M19413439+4017482 & NGC 6819 & 15641.7, 16121.4, 16323.4, 16530.0, 16744.7 & 13 $\pm$ 2 & 13 $\pm$ 1 \\
2M19412658+4011418 & NGC 6819 & 15641.7, 16323.4, 16530.0 & 8 $\pm$ 2 & 10 $\pm$ 2 \\
2M23571013+5647167 & NGC 7789 & 16327.3 & 3 $\pm$ 0.7 & $>$6 \\
2M23570744+5641417 & NGC 7789 & 15641.7, 16121.4,  16530.0, 16744.7 & 15 $\pm$ 1 & 17 $\pm$ 3 \\
2M23555312+5641203 & NGC 7789 & 15641.7, 16121.4, 16323.4, 16326.0, 16530.0 & 9 $\pm$ 1 & 11 $\pm$ 3 \\
2M23581471+5651466 & NGC 7789 & 15641.7, 16121.4, 16530.0 & 9 $\pm$ 0.9 & 10 $\pm$ 0.9 \\
2M12400451-6036566 & Trumpler 20 & 15641.7, 16121.4, 16530.0 & 14 $\pm$ 4 & 15 $\pm$ 2 \\
2M12390411-6034001 & Trumpler 20 & 15641.7, 16323.4, 16326.0, 16530.0 & 10 $\pm$ 2 & 10 $\pm$ 1 \\
2M12402480-6043101 & Trumpler 20 & 15641.7, 16326.0 & 10 $\pm$ 3 & $>$10 $\pm$ 0.6 \\
2M12400755-6035445 & Trumpler 20 & 15641.7, 16121.4, 16323.4, 16530.0 & 10 $\pm$ 0.9 & 10 $\pm$ 0.5 \\
2M12400260-6039545 & Trumpler 20 & 15641.7, 16323.4, 16530.0 & 9 $\pm$ 2 & 10 $\pm$ 2 \\
2M12391003-6038402 & Trumpler 20 & 15641.7, 16121.4, 16323.4, 16326.0, 16530.0 & 8 $\pm$ 3 & 11 $\pm$ 2 \\
2M12402949-6038518 & Trumpler 20 & 15641.7, 16326.0, 16530.0 & 8 $\pm$ 3 & 11 $\pm$ 1 \\
2M12385807-6030286 & Trumpler 20 & 15641.7, 16121.4, 16530.0 & 6 $\pm$ 3 & 8 $\pm$ 2 \\
2M12392699-6036052 & Trumpler 20 & 15641.7 & 6 $\pm$ 1 & $>$9 \\ \hline
\end{tabular}
\end{table*}

\begin{figure*}
\centering
\includegraphics[scale=0.65]{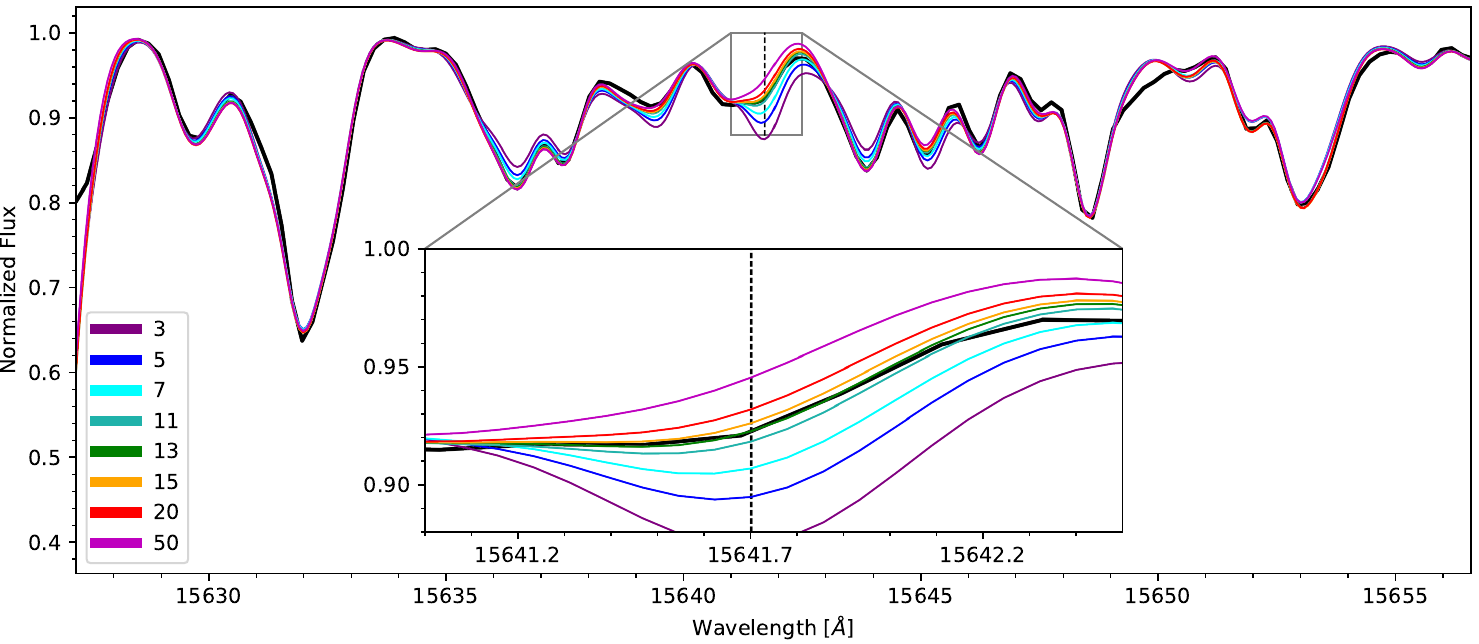}
\includegraphics[scale=0.65]{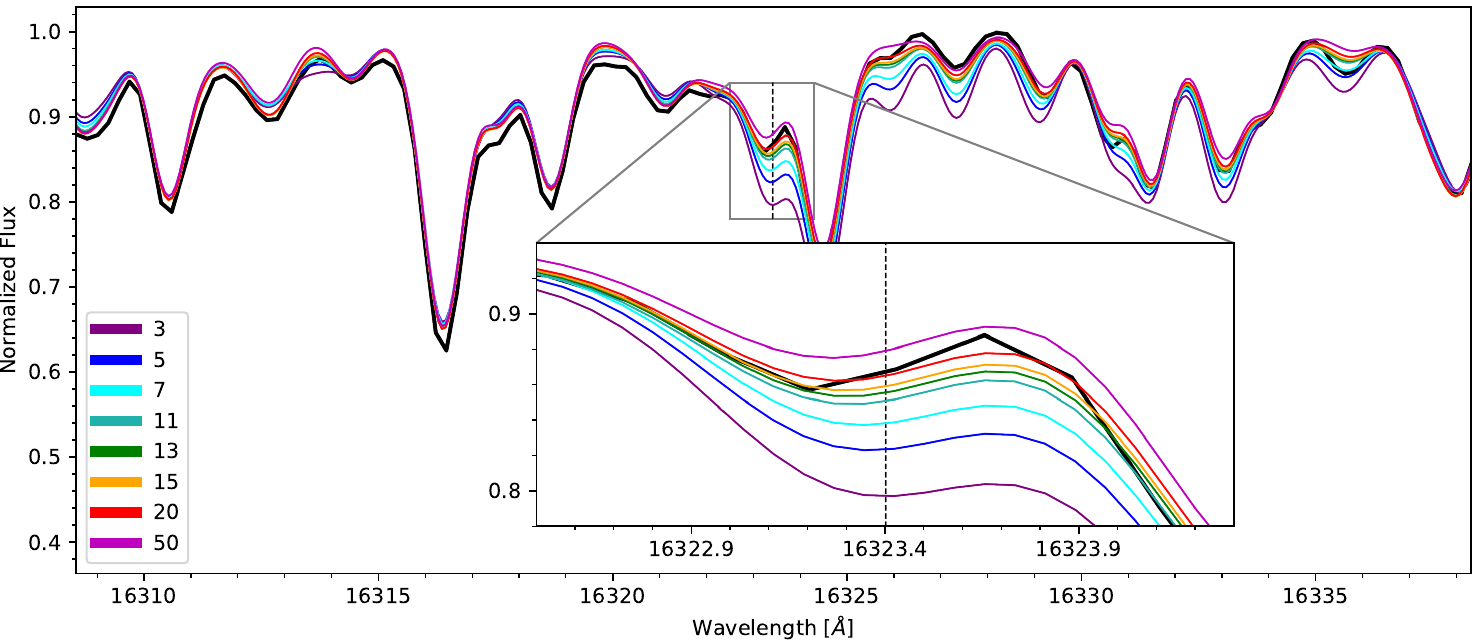}
\includegraphics[scale=0.65]{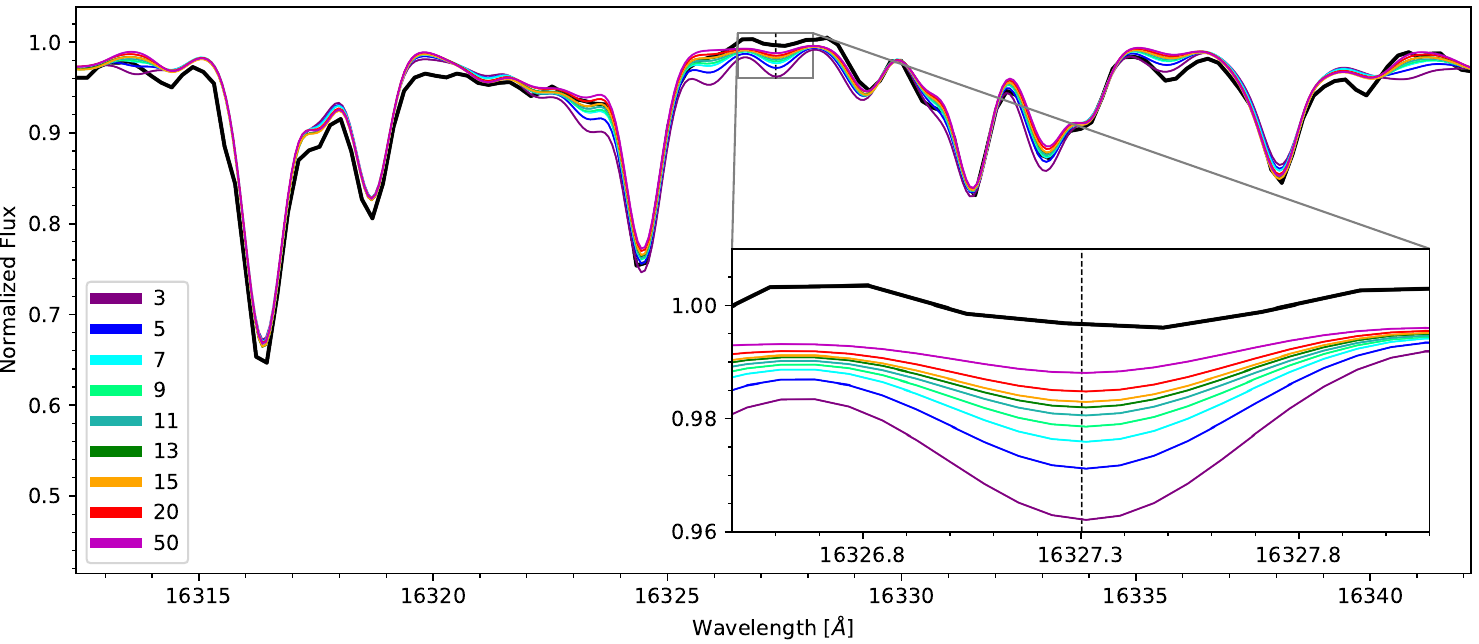}
\caption{Examples of the models used by BACCHUS to fit spectral lines and derive \cratio ratios. Each {\it colored line} represents a different value for \cratio. The {\it black line} is the observed spectrum, and the vertical, {\it dashed line} marks the central wavelength of each spectral feature. \textit{Top}: An example of a well-fit line characterized as a measurement where the dark green, \cratio$=13$ model provides the best fit for this line (star: 2M19413439+4017482; cluster: NGC 6819; average stellar \cratio$=13$). \textit{Middle}: An example of a line characterized as a limit where the \cratio$=15$ ({\it orange}) and $20$ ({\it red}) models provide the closest fits and slightly overestimate this star's \cratio (star: 2M19413439+4017482; cluster: NGC 6819; average stellar \cratio$=13$). \textit{Bottom}: An example of a poor fit characterized as a non-measurement where the models clearly miss the spectrum and estimate an unreasonably large value ($>$50) for this particular line (star: 2M00571844+8510288; cluster: NGC 188; average stellar \cratio$=13$).}
\label{fig:fits}
\end{figure*}

\noindent The \cratio ratios reported in the BAWLAS VAC were determined using the BACCHUS code to fit CO and CN lines in eight windows centered on 15641.7 \AA, 16121.4 \AA, 16323.4 \AA, 16326.0 \AA, 16327.3 \AA, 16530.0 \AA, 16741.2 \AA, and 16744.7 \AA. This approach allows for the efficient processing of such a large data set, however, there is always the possibility that some spectra are affected by noise and/or have poorly fit features. To ensure that our observed \cratio-age/mass relations are accurate, we visually inspected all eight spectral fits for all 49 stars in our sample and manually vetted the \cratio ratios, updating the values as necessary.

For each star, we characterized the fit to the CO or CN line in each of the eight spectral windows as a ``measurement" (i.e., the line is acceptably well fit), a ``limit" (i.e., the line is decently fit but could be better), or a ``non-measurement" (i.e., the fit is not representative of the observed spectrum). These categories were assigned after several inspections of each spectral fit since defining what is a ``good'' fit versus a ``bad'' fit can be somewhat subjective. Factors such as (1) how well the synthetic spectra matched the shape of the observed spectra and (2) whether the synthetic spectra were noticeably shifted above or below the observed spectral feature were considered in this process. Both of these factors could act to artificially increase or decrease the derived \cratio, so careful consideration was given to identify these biases. Examples of measurement, limit, and non-measurement spectral fits are shown in Figure \ref{fig:fits}.

Once the fits to each star's eight spectral features were characterized as well fit or not, we computed the final \cratio ratios for each star. The BACCHUS code derives a separate \cratio value based on the fit for each of the eight spectral features, so we computed a given star's total \cratio by averaging the ratio values provided for each well-fit line. All spectral fits falling into the ``measurement" or ``limit" categories, such as the top and middle panels in Figure \ref{fig:fits}, were used to determine the star's final \cratio value. We report the standard deviation of these \cratio values from well-fit spectral lines as the \cratio error. We note that this error calculation often underestimates the true error, especially for stars with fewer well-fit lines that produce a measurement, and it does not take into account systematic errors in the measurement and modelling processes (see \citealt{HAYES}).

Overall, we found only three instances of stars that had generally poor fits for most of their eight spectral features. The APOGEE IDs (and clusters) for these three stars are 2M19212437+3735402 (NGC 6791), 2M23571013+5647167 (NGC 7789), and 2M12392699-6036052 (Trumpler 20), and their \cratio ratios are reported as \cratio lower limits.

While any combination of the eight spectral features could give the final \cratio value,  we required that each star must have well fit 15641.7 \AA\ and 16530.0 \AA\ lines as a means to bring some level of standardization to the process.  These lines were chosen because they were the most common lines with generally good fits in our sample, and stars displaying generally questionable fits were often lacking good fits for at least one of these two lines. Stars that display poor fits for at least one of these two lines have their \cratio value shown as a lower limit. The following three stars were excluded from the sample after imposing this condition: 2M20190397+3745002 (Berkeley 85), 2M19213635+3739445 (NGC 6791), and 2M12402480-6043101 (Trumpler 20). Adopting this last criterion brings our final sample to 43 stars with \cratio ratio measurements and six stars providing \cratio limits. We adopt the manually vetted \cratio values for the subsequent analysis in this paper.

The results of this spectral analysis can be seen in Table \ref{tab:stars}, which gives the stellar parameters, our manually vetted \cratio values, OCCAM cluster membership probabilities, and evolutionary states (RGB or red clump) for all stars that were determined to have reliable \cratio ratios. Table \ref{tab:C12C13_initial_final} includes the final sample stars with \cratio measurements and those stars with \cratio limits. This table displays the BAWLAS VAC \cratio values as well as our manually determined \cratio values for easy comparison between the two analyses. Additionally, the table lists the lines we used in determining the ratio for each star.

\section{Results}
\label{sec:results}

\subsection{Evolution of \cratio with $\log{g}$}

\begin{figure*}
\centering
\includegraphics[scale=1.2]{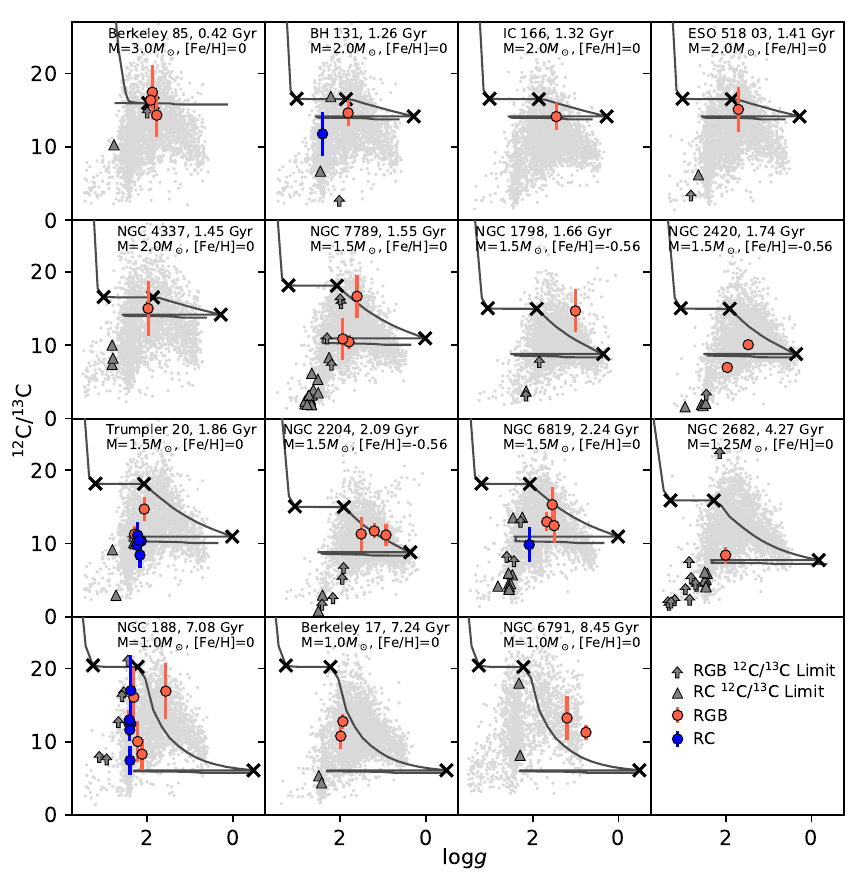}
\caption{The evolution of the \cratio ratio as a function of $\log{g}$ for each cluster. Typical errors in $\log{g}$ are 0.02 dex. {\it Blue circles} correspond to red clump (core He-burning) stars, and {\it orange circles} correspond to stars on the RGB. The Lagarde models are shown as {\it dark gray curves}, and the three {\it black $\times$} symbols mark the end of the first dredge-up, the RGB luminosity bump, and the tip of the RGB (left to right) on each model. {\it Light gray points} represent field stars in the BAWLAS VAC that have \cratio and [Fe/H] within 0.03 dex of each cluster mean [Fe/H]; we note that these field stars are not necessarily the same age as the cluster stars. {\it Dark gray arrows} are \cratio limit stars belonging to each cluster.}
\label{fig:12C13C_logg}
\end{figure*}

\noindent A crucial test for understanding the nature of extra mixing on the RGB is to observe how the \cratio ratio evolves with time, or equivalently surface gravity ($\log{g}$), on the RGB and red clump (RC) and compare this evolution to models that take into account an extra mixing mechanism. Figure \ref{fig:12C13C_logg} presents the \cratio evolution with $\log{g}$ for our open cluster stars (orange and blue circles), separated into each cluster. Additionally, the \cratio limit stars determined in the BAWLAS VAC and through our manual spectral fit examination are shown as dark gray arrows, and the light gray points represent stars in the BAWLAS VAC with [Fe/H] within 0.03 dex of the cluster mean [Fe/H]. In Figure \ref{fig:12C13C_logg}, we show models from \citet{LAGARDE12} (hereafter referred to as the ``Lagarde models''; dark gray, solid curves) that exhibit extra mixing effects caused by the combination of thermohaline instability and stellar rotation. Stars with mass above $\backsim$2.2\msun at near-solar metallicity are not expected to reach the luminosity bump (e.g., \citealt{CL10,LAGARDE19}), so the model representing more massive stars (i.e., Berkeley 85) exhibit extra mixing effects due to stellar rotation rather than thermohaline instability. Less massive models exhibit a combined effect, but the thermohaline instability dominates extra mixing (e.g., \citealt{CL10}).

The Lagarde models were generated for discrete mass ($M$) and metallicity ([Fe/H]) values ranging from $M$ = 0.85 to 6 \msun and [Fe/H] = -2.16 to 0. For comparison with the data, we chose the model with the closest mass and [Fe/H] values to each cluster (see Table \ref{tab:clusters} for the average, APOGEE-measured [Fe/H] and initial stellar mass from the MIST isochrone for each cluster). In Figure \ref{fig:12C13C_logg}, the Lagarde models encompass the time just after the end of the first dredge-up (the leftmost black $\times$ symbol in each subplot at relatively high $\log{g}$ and high \cratio) until the early AGB (the lower horizontal sections at relatively low $\log{g}$ and low \cratio). Notably, the RGB luminosity bump (the middle $\times$ in each subplot) is clearly seen in most of the models at the point where there is a sudden, large decrease in \cratio (generally at $\log{g}$ $\approx$ 1.5--2). Before this point, there are small changes in the surface \cratio ratio due to rotation, however, after this point the HBS comes into contact with the envelope and thermohaline extra mixing alters the surface \cratio ratio much more dramatically. Following the extra mixing dip, the models begin to flatten again just after the tip of the RGB (rightmost $\times$ in each subplot in Figure \ref{fig:12C13C_logg}) as the star begins core helium burning on the horizontal branch.

In Figure \ref{fig:12C13C_logg}, the open cluster stars in our sample have been differentiated by color to show stars at two evolutionary stages: orange circles represent stars on the RGB, either just before or currently undergoing extra mixing, and blue circles represent stars that have finished extra mixing on the RGB and are RC stars. The \cratio limit stars are also classified as RGB or RC, as indicated by the shape of the gray arrow. We utilized the ``SPEC\_RGB'' and ``SPEC\_RC'' flags in the APOGEE ``PARAMFLAG'' array to classify stellar evolutionary states in our sample. Originally, these flags were populated based on the work of \citet{JONSSON20} which separated the stars in APOGEE based on evolutionary state to more accurately calibrate surface gravities for similar stars. \citet{JONSSON20} categorized the APOGEE stars as dwarf, RGB, or RC based on a given star's spectroscopic $\log{g}$ and $T_{eff}$, total metallicity ([M/H]), and [C/N] values falling within a range typical of each evolutionary group; see \S5.2.2 in \citet{JONSSON20} for the specific values used to determine each group. We note that due to the difficult nature of distinguishing between the early RGB and the RC, there are potential misclassifications. Therefore, these assigned evolutionary states should be taken as an approximation. Table \ref{tab:stars} lists the evolutionary state for each star in our sample.

Though data for each cluster are sparse compared to the range of each Lagarde model shown, in general we see that the data line up with the models within the margin of error. Also in nearly every case, the RGB stars have higher \cratio ratios than the clump stars, which is expected because clump stars have fully completed extra mixing on the RGB while RGB stars have not.

\subsection{\cratio as a Function of Age and Mass}
\label{subsec:agemass}

\begin{figure*}
\centering
\includegraphics[scale=0.4]{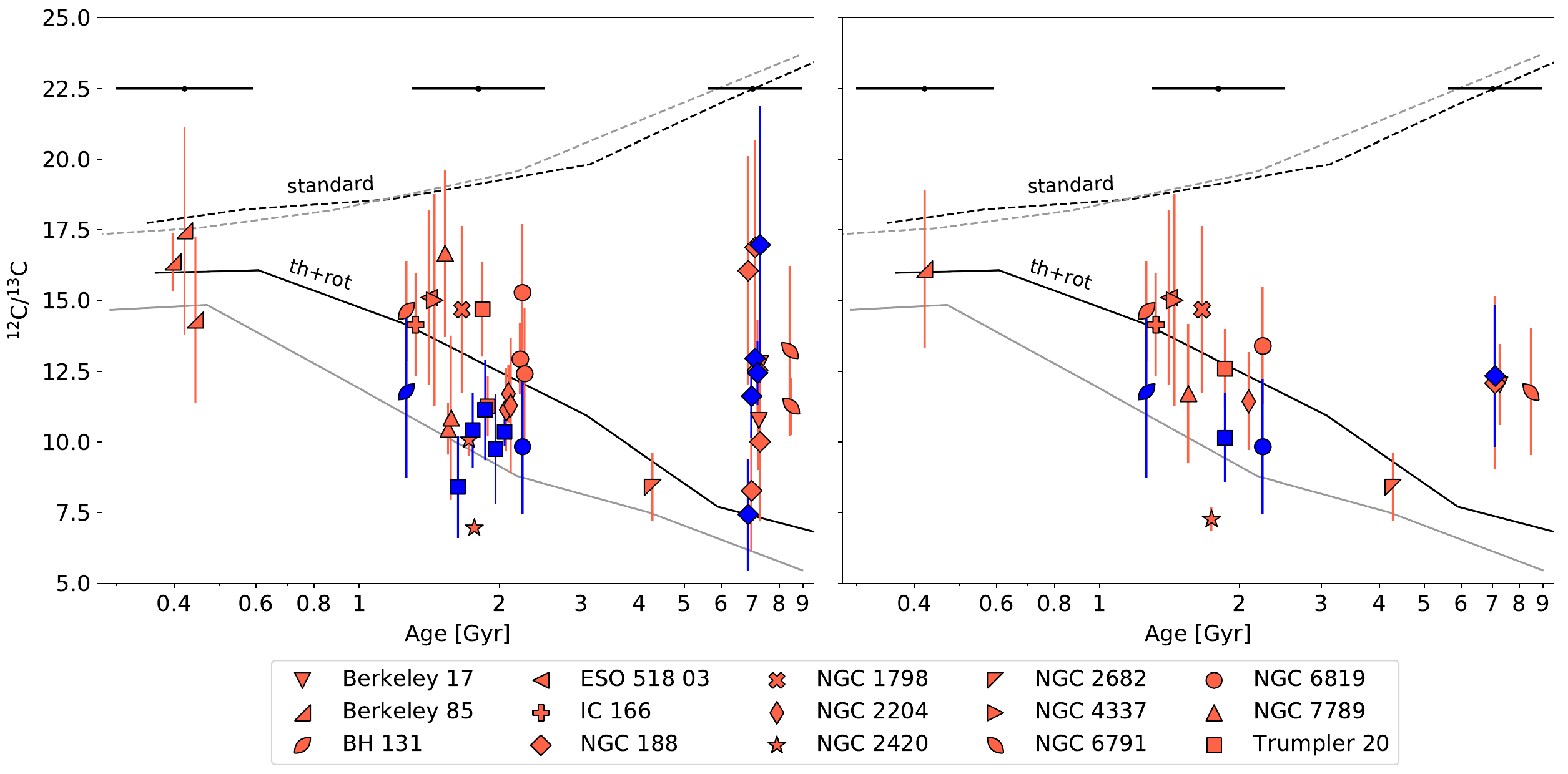}
\caption{The \cratio ratio as a function of age for individual open cluster stars ({\it left}) and mean cluster values ({\it right}). {\it Blue data points} correspond to red clump stars undergoing core He burning, and {\it orange data points} correspond to stars on the RGB. Individual stars in a given cluster, or at the same age, are intentionally offset slightly in age to better show the \cratio error bars. Horizontal, {\it black error bars} at the top of each figure represent typical age errors. Overlaid are curves showing the \cratio and age values at the tip of the RGB from the Lagarde models assuming (1) standard stellar evolution ({\it dashed curves}) and (2) thermohaline mixing and stellar rotation ({\it solid curves}) for [Fe/H] = 0 ({\it black}) and [Fe/H] = -0.56 ({\it gray}).}
\label{fig:12C13C_age}
\end{figure*}

\begin{figure*}
\centering
\includegraphics[scale=0.4]{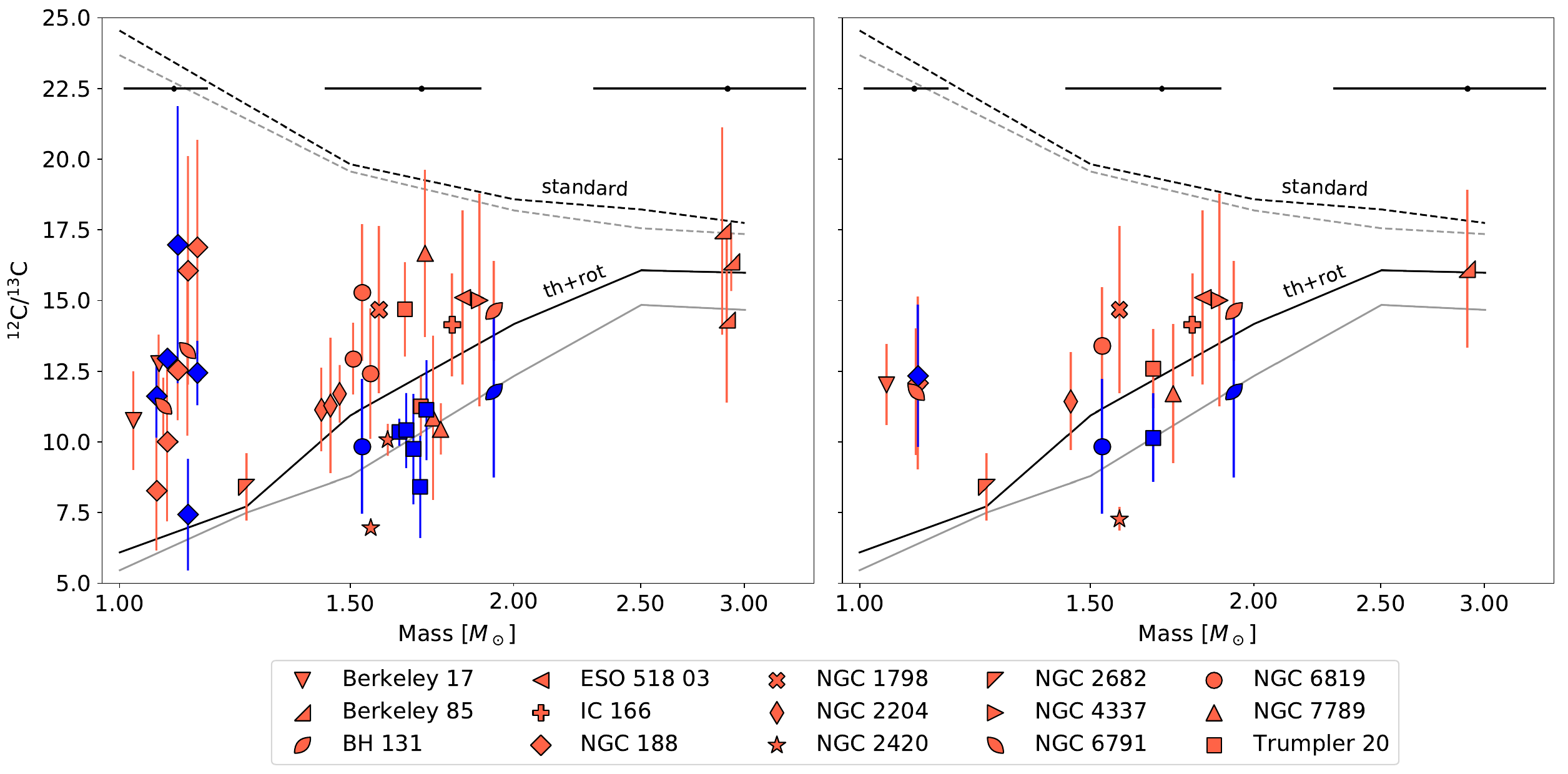}
\caption{The same as Figure \ref{fig:12C13C_age}, but as a function of the initial stellar mass for stars at the cluster main sequence turn off instead of stellar age. Horizontal, {\it black error bars} at the top of each figure represent typical mass errors.}
\label{fig:12C13C_mass}
\end{figure*}

\noindent We are also interested in how the \cratio ratio and extra mixing change for stars of different ages or masses. The \cratio ratios as a function of age and mass for our sample are presented in Figures \ref{fig:12C13C_age} and \ref{fig:12C13C_mass} respectively. Both figures display the \cratio ratios from each individual star (left) and the mean \cratio value for each cluster and evolutionary group (right), using the same evolutionary state color conventions as previously described. We note that in the individual star measurement plots, the data points for a given cluster are offset in age or mass randomly, covering a small range around the true cluster age or mass to allow for better visualization of the error bars for a given star. 

As shown in Figure \ref{fig:12C13C_age}, the \cratio scatter at a given age is reduced when one accounts for the stellar evolutionary state (see left panel of Figure \ref{fig:12C13C_age}). For example, in the 1--2.5 Gyr range, two distinct groupings show the same decreasing trend in \cratio with increasing age; however, the less evolved stars (shown as orange symbols) that are most likely still involved in thermohaline extra mixing on the RGB have slightly higher \cratio ratios than the more evolved stars (blue symbols). Likewise, the \cratio-mass relation exhibits a similar split between evolutionary groups in the 1.5--2 \msun range with the less evolved RGB stars exhibiting the same trend with stellar mass as the more evolved stars, just offset to higher \cratio ratios.

Overlaying the data in Figures \ref{fig:12C13C_age} and \ref{fig:12C13C_mass} are models, again from \citet{LAGARDE12}, that show the predicted \cratio-age/mass trend for both the standard mixing theory (i.e., only convection; dashed curve) and one that includes extra mixing due to the thermohaline instability and stellar rotation (solid curve). The colors of the models indicate metallicity, where black is for [Fe/H] = 0 and gray is for [Fe/H] = -0.56. The models shown are the predicted \cratio values for stars at the tip of the RGB. Therefore, if stars with mass less than $\backsim$2.2\msun experience thermohaline extra mixing, we expect the orange points to fall slightly above the extra mixing model since these stars have not been mixed as much as the model; furthermore, the blue points should fall at or slightly below the extra mixing model, since these stars have undergone all of RGB thermohaline extra mixing but could be slightly more mixed due to rotation.

We find that our data exhibit \cratio values consistently lower than the standard mixing model predictions and have slight trends with age and mass, which implies a need for extra mixing on the RGB to explain these observations. The data agree more closely with the thermohaline and rotation model, though there are notable discrepancies, specifically toward the older/less massive regime, that require further investigation and could provide useful information to improve our extra mixing understanding.

\section{Discussion}
\label{sec:discussion}

We have studied how the \cratio ratio changes for red giants of varying evolutionary state and age/mass. Because these characteristics each affect extra mixing efficiency, drawing conclusions about the entire sample as a whole is difficult. Instead, we discuss here the \cratio trends and uncertainties within subgroups of stars in our sample with similar characteristics.

First, the stars belonging to Berkeley 85 are predicted to have an initial mass of $2.91^{+0.43}_{-0.61}$\msun and have slightly super solar metallicity ([Fe/H] = 0.10 $\pm$ 0.01). Stars with these parameters are not thought to reach the conditions for thermohaline mixing to occur since the non-degenerate He core begins He burning before the RGB bump can be reached (e.g., \citealt{CL10,LAGARDE19}). Therefore, the predominant mechanism for mixing in these stars is stellar rotation. The effect of stellar rotation slightly lowering the \cratio on the RGB can be seen by comparing the standard model and the extra mixing model predictions at this age/mass in Figures \ref{fig:12C13C_age} and \ref{fig:12C13C_mass}.

Next, stars with initial mass below $\backsim$2.2\msun (e.g., \citealt{CL10,LAGARDE19}) are expected to experience the conditions for thermohaline extra mixing. Moreover, thermohaline extra mixing is thought to be the dominating extra mixing mechanism for these stars on the RGB. In the 1--2.5 Gyr, or 1.5--2 \msun, range in Figures \ref{fig:12C13C_age} and \ref{fig:12C13C_mass}, we observe generally good agreement with the models. That is, most of the orange stars have \cratio values higher than the model, since these stars have not undergone as much mixing as stars in the model have at the tip of the RGB, and the blue stars have \cratio values at or lower than the model given \cratio error bars since these stars have undergone the full extent of RGB extra mixing.

Notably, in the 1--2.5 Gyr/1.5--2 \msun\ range, there are three clusters that have subsolar metallicities that fall closer to the gray model (i.e. [Fe/H] = -0.56): NGC 1798, NGC 2420, and NGC 2204. Specifically for NGC 2420, these stars tend to have \cratio values that fall more in line with the gray model. The lower metallicity, gray model predicts lower \cratio values than the black model implying more mixing has occurred for these more metal-poor stars. However, from these data, it is uncertain if this is a general trend attributed to the metallicity of the star since the stars in NGC 1798 and NGC 2204 have \cratio values that are relatively similar to the near-solar metallicity stars and the black model. Additionally, there is always the question whether the stars in NGC 2420 are actually RC stars that have undergone the full extent of RGB extra mixing, which could explain the lower \cratio values.

Finally, the clusters in the old, low-mass range (less than $\backsim$1.25\msun) exhibit notable deviations from the thermohaline model. Most of these stars show relatively high \cratio values compared to the extra mixing model. Though unknown at present, these differences could possibly be due to (1) systematics in our \cratio measurements, (2) the need for more careful analysis of the evolutionary states, (3) the need for fine tuning of the extra mixing models so that they better match observations, or (4) some combination of these factors.

One obvious cluster that deviates from the thermohaline model in Figures \ref{fig:12C13C_age} and \ref{fig:12C13C_mass} is NGC 188. This cluster exhibits relatively high \cratio ratios for nearly all of its stars compared to the thermohaline model. Additionally, there are a few stars belonging to the cluster that have non-intuitive evolutionary state classifications and \cratio ratios. Notably, the two RGB (orange) stars with low \cratio ($\backsim$10) at $\log{g} \backsim$ 2 (see Figure \ref{fig:12C13C_logg}) are peculiar because they should not have experienced the luminosity bump and extra mixing given their $\log{g}$ values if they are truly on the RGB, and yet they have such low \cratio ratios. Similarly, one star classified as a RC star shows \cratio $\backsim$ 17, which is abnormally high for a star that should have fully undergone extra mixing.

We can only conjecture possible explanations for these anomalies. Perhaps these peculiar stars have been misclassified as NGC 188 members, or they are members misclassified as RGB or RC and are actually in some other phase of evolution, such as the AGB. Alternatively, there could be some missing dependency of extra mixing, such as a spread in stellar rotation speeds and mixing efficiencies, that we are not directly considering that could provide an explanation for these unexpected results. NGC 188 presents itself as an interesting case study as it provided the largest number (10) of stars with reliably determined \cratio ratios for a single cluster in this study, and the cluster seems to show a large intrinsic spread in these \cratio ratios for both the RGB and RC evolutionary states that are not well explained by either the standard or thermohaline model.

The data presented here cover a large range of stellar characteristics which affect RGB extra mixing, making it difficult to attribute just one model to explain all of the \cratio observations. The Lagarde thermohaline and rotation models do a sufficient job for some of the data. There have been concerns raised in the literature, however, about modelling the thermohaline instability to explain RGB extra mixing in general and if this instability is physically able to cause a large-scale change in surface abundances (e.g., see \citealt{TAYAR22} and references therein). Most notably, hydrodynamical simulations (e.g., \citealt{DENISSENKOV10}, \citealt{DENISSENKOV11}, \citealt{TRAXLER11}) have yielded short, wide ``salt fingers" that transport material, meaning the so-called ``C'' parameter, which is related to the ratio of the salt finger's length to diameter, is on the order of 1 to a few (\citealt{KIPPENHAHN80}). For the thermohaline instability to reach the convective envelope of the star, one needs to incorporate long, thin salt fingers; often a C value on the order of 1000 is adopted to ensure the instigation of these long fingers and extra mixing with the envelope (e.g., \citealt{ULRICH72}, \citealt{CZ07}). This large difference in C causes some to question our understanding of the thermohaline instability in RGB stars and its ability to explain extra mixing.

Incorporating multiple physical processes, such as thermohaline instability and stellar rotation as in the case of the Lagarde models, is one way that authors have been able to fine tune the models to produce the observed abundances. However, like some other studies attempting to model extra mixing, the Lagarde models have treated the thermohaline mixing and rotation-induced mixing independently and then simply added their effects, whereas these two processes may well interact in a real system, affecting how each process evolves. Studies incorporating interacting extra mixing processes are not widely available yet, though there have been attempts thus far (e.g., \citealt{MAEDER13, SG18}). In the future, these models will ideally shed more light onto the mixing conditions in red giants.

\section{Conclusions}
\label{sec:conclusions}

\noindent We have investigated non-canonical, extra mixing in red giant stars by observing how the \cratio ratio, a tracer of red giant internal mixing, evolves on the RGB, and we have studied how this chemical evolution varies with stellar age and mass. To do so, we first isolated a sample of 43 red giant stars, reliably identified as belonging to one of 15 surveyed open clusters, from the APOGEE DR17 data set that have \cratio ratios derived as a part of the BAWLAS VAC \citep{HAYES}. We then identified the evolutionary state of each star using the APOGEE RGB/RC flags. Finally, to test the overall importance of extra mixing in predicting red giant surface abundances and to gain insight into its instigator, we compared how the \cratio ratios varied with evolutionary state, age, and mass to how they are expected to vary based on two theoretical models: (1) standard stellar evolution (i.e., no extra mixing on the RGB) and (2) RGB extra mixing \citep[in this case, we used the thermohaline and stellar rotation model from][]{LAGARDE12}. Our results are highlighted in Figures \ref{fig:12C13C_logg}, \ref{fig:12C13C_age}, and \ref{fig:12C13C_mass}.

While the details of thermohaline mixing are still debated, our data set of red giant stars with varying mass, evolutionary state, and homogeneously derived \cratio show a clear need for extra mixing of some form along the RGB. We find that the \cratio in stars with mass greater than $\backsim$2.5\msun can be explained by stellar rotation since thermohaline extra mixing is not expected to occur in these stars. Additionally, we find that the Lagarde models do a reasonable job of matching the general trends exhibited by the observations and suggest that the source of extra mixing must produce similar trends. Specifically, \cratio tends to decrease with increasing age or decreasing mass in the age/mass range 1-2 Gyr/1.5-2\msun. From these data, the \cratio likely decreases more, or mixing is more efficient, for lower metallicity stars at a given age/mass in this range. Finally, stars in our data set less massive than $\backsim$1.25\msun\ tend to deviate from model predictions, so either additional, detailed observations and analyses are needed to justify this trend or the thermohaline model prescription needs to be revised to explain these observed mixing indicators. The growing availability of similar high quality data will ultimately help constrain the physics of extra mixing and inform how to accurately model what is happening in stellar interiors during these dramatic events.

\section*{Acknowledgements}

CM, SRM, and AA acknowledge support from National Science Foundation grant AST-1909497.

TM acknowledges financial support from the Spanish Ministry of Science and Innovation (MICINN) through the Spanish State Research
Agency, under the Severo Ochoa Program 2020-2023 (CEX2019-000920-S) as well as support from the ACIISI, Consejería de Economía, Conocimiento y Empleo del Gobierno de Canarias and the European Regional Development Fund (ERDF) under grant with reference PROID2021010128.

Funding for the Sloan Digital Sky Survey IV has been provided by the Alfred P. Sloan Foundation, the U.S. Department of Energy Office of Science, and the Participating Institutions. 

SDSS-IV acknowledges support and resources from the Center for High Performance Computing at the 
University of Utah. The SDSS website is www.sdss.org.

SDSS-IV is managed by the Astrophysical Research Consortium for the Participating Institutions 
of the SDSS Collaboration including the Brazilian Participation Group, the Carnegie Institution for Science, Carnegie Mellon University, Center for Astrophysics | Harvard \& Smithsonian, the Chilean Participation Group, the French Participation Group, Instituto de Astrof\'isica de Canarias, The Johns Hopkins University, Kavli Institute for the Physics and Mathematics of the Universe (IPMU) / University of Tokyo, the Korean Participation Group, Lawrence Berkeley National Laboratory, Leibniz Institut f\"ur Astrophysik Potsdam (AIP),  Max-Planck-Institut f\"ur Astronomie (MPIA Heidelberg), Max-Planck-Institut f\"ur Astrophysik (MPA Garching), Max-Planck-Institut f\"ur Extraterrestrische Physik (MPE), National Astronomical Observatories of China, New Mexico State University, New York University, University of Notre Dame, Observat\'ario Nacional / MCTI, The Ohio State University, Pennsylvania State University, Shanghai Astronomical Observatory, United Kingdom Participation Group, Universidad Nacional Aut\'onoma de M\'exico, University of Arizona, University of Colorado Boulder, University of Oxford, University of Portsmouth, University of Utah, University of Virginia, University of Washington, University of Wisconsin, Vanderbilt University, and Yale University.

This is a pre-copyedited, author-produced PDF of an article accepted for publication in Monthly Notices of the Royal Astronomical Society following peer review. The version of record is available online at: \href{https://doi.org/10.1093/mnras/stad2156}{https://doi.org/10.1093/mnras/stad2156}.

%%%%%%%%%%%%%%%%%%%%%%%%%%%%%%%%%%%%%%%%%%%%%%%%%%
\section*{Data Availability}

The data underlying this article were accessed from \href{https://www.sdss.org/dr17/irspec/spectro_data/}{SDSS DR17}. The derived data generated in this research will be shared on reasonable request to the corresponding author.

%%%%%%%%%%%%%%%%%%%% REFERENCES %%%%%%%%%%%%%%%%%%

\bibliographystyle{mnras}
\bibliography{references, software_references}

\begin{thebibliography}{}
\makeatletter
\relax
\def\mn@urlcharsother{\let\do\@makeother \do\$\do\&\do\#\do\^\do\_\do\%\do\~}
\def\mn@doi{\begingroup\mn@urlcharsother \@ifnextchar [ {\mn@doi@}
  {\mn@doi@[]}}
\def\mn@doi@[#1]#2{\def\@tempa{#1}\ifx\@tempa\@empty \href
  {http://dx.doi.org/#2} {doi:#2}\else \href {http://dx.doi.org/#2} {#1}\fi
  \endgroup}
\def\mn@eprint#1#2{\mn@eprint@#1:#2::\@nil}
\def\mn@eprint@arXiv#1{\href {http://arxiv.org/abs/#1} {{\tt arXiv:#1}}}
\def\mn@eprint@dblp#1{\href {http://dblp.uni-trier.de/rec/bibtex/#1.xml}
  {dblp:#1}}
\def\mn@eprint@#1:#2:#3:#4\@nil{\def\@tempa {#1}\def\@tempb {#2}\def\@tempc
  {#3}\ifx \@tempc \@empty \let \@tempc \@tempb \let \@tempb \@tempa \fi \ifx
  \@tempb \@empty \def\@tempb {arXiv}\fi \@ifundefined
  {mn@eprint@\@tempb}{\@tempb:\@tempc}{\expandafter \expandafter \csname
  mn@eprint@\@tempb\endcsname \expandafter{\@tempc}}}

\bibitem[\protect\citeauthoryear{{Abdurro'uf} et~al.,}{{Abdurro'uf}
  et~al.}{2022}]{ABDURROUF22}
{Abdurro'uf} et~al., 2022, \mn@doi [\apjs] {10.3847/1538-4365/ac4414}, \href
  {https://ui.adsabs.harvard.edu/abs/2022ApJS..259...35A} {259, 35}

\bibitem[\protect\citeauthoryear{{Allende Prieto}, {Beers}, {Wilhelm},
  {Newberg}, {Rockosi}, {Yanny}  \& {Lee}}{{Allende Prieto}
  et~al.}{2006}]{AllendePrieto2006}
{Allende Prieto} C.,  {Beers} T.~C.,  {Wilhelm} R.,  {Newberg} H.~J.,
  {Rockosi} C.~M.,  {Yanny} B.,   {Lee} Y.~S.,  2006, \mn@doi [\apj]
  {10.1086/498131}, \href {http://adsabs.harvard.edu/abs/2006ApJ...636..804A}
  {636, 804}

\bibitem[\protect\citeauthoryear{{Beaton} et~al.,}{{Beaton}
  et~al.}{2021}]{Beaton2021}
{Beaton} R.~L.,  et~al., 2021, \mn@doi [\aj] {10.3847/1538-3881/ac260c}, \href
  {https://ui.adsabs.harvard.edu/abs/2021AJ....162..302B} {162, 302}

\bibitem[\protect\citeauthoryear{{Blanton} et~al.,}{{Blanton}
  et~al.}{2017}]{BLANTON17}
{Blanton} M.~R.,  et~al., 2017, \mn@doi [\aj] {10.3847/1538-3881/aa7567}, \href
  {https://ui.adsabs.harvard.edu/abs/2017AJ....154...28B} {154, 28}

\bibitem[\protect\citeauthoryear{{Boothroyd} \& {Sackmann}}{{Boothroyd} \&
  {Sackmann}}{1999}]{BS99}
{Boothroyd} A.~I.,  {Sackmann} I.~J.,  1999, \mn@doi [\apj] {10.1086/306546},
  \href {https://ui.adsabs.harvard.edu/abs/1999ApJ...510..232B} {510, 232}

\bibitem[\protect\citeauthoryear{{Boothroyd}, {Sackmann}  \&
  {Wasserburg}}{{Boothroyd} et~al.}{1995}]{BSW95}
{Boothroyd} A.~I.,  {Sackmann} I.~J.,   {Wasserburg} G.~J.,  1995, \mn@doi
  [\apjl] {10.1086/187806}, \href
  {https://ui.adsabs.harvard.edu/abs/1995ApJ...442L..21B} {442, L21}

\bibitem[\protect\citeauthoryear{{Bossini} et~al.,}{{Bossini}
  et~al.}{2019}]{BOSSINI19}
{Bossini} D.,  et~al., 2019, \mn@doi [\aap] {10.1051/0004-6361/201834693},
  \href {https://ui.adsabs.harvard.edu/abs/2019A&A...623A.108B} {623, A108}

\bibitem[\protect\citeauthoryear{{Bowen} \& {Vaughan}}{{Bowen} \&
  {Vaughan}}{1973}]{bv73}
{Bowen} I.~S.,  {Vaughan} A.~H. J.,  1973, \mn@doi [\ao]
  {10.1364/AO.12.001430}, \href
  {https://ui.adsabs.harvard.edu/abs/1973ApOpt..12.1430B} {12, 1430}

\bibitem[\protect\citeauthoryear{{Brogaard} et~al.,}{{Brogaard}
  et~al.}{2021}]{BROGAARD21}
{Brogaard} K.,  et~al., 2021, \mn@doi [\aap] {10.1051/0004-6361/202140911},
  \href {https://ui.adsabs.harvard.edu/abs/2021A&A...649A.178B} {649, A178}

\bibitem[\protect\citeauthoryear{{Busso}, {Wasserburg}, {Nollett}  \&
  {Calandra}}{{Busso} et~al.}{2007}]{BUSSO07}
{Busso} M.,  {Wasserburg} G.~J.,  {Nollett} K.~M.,   {Calandra} A.,  2007,
  \mn@doi [\apj] {10.1086/522616}, \href
  {https://ui.adsabs.harvard.edu/abs/2007ApJ...671..802B} {671, 802}

\bibitem[\protect\citeauthoryear{{Cantat-Gaudin} et~al.,}{{Cantat-Gaudin}
  et~al.}{2020}]{CG20}
{Cantat-Gaudin} T.,  et~al., 2020, \mn@doi [\aap]
  {10.1051/0004-6361/202038192}, \href
  {https://ui.adsabs.harvard.edu/abs/2020A&A...640A...1C} {640, A1}

\bibitem[\protect\citeauthoryear{{Chanam{\'e}}, {Pinsonneault}  \&
  {Terndrup}}{{Chanam{\'e}} et~al.}{2005}]{CHANAME05}
{Chanam{\'e}} J.,  {Pinsonneault} M.,   {Terndrup} D.~M.,  2005, \mn@doi [\apj]
  {10.1086/432410}, \href
  {https://ui.adsabs.harvard.edu/abs/2005ApJ...631..540C} {631, 540}

\bibitem[\protect\citeauthoryear{{Charbonnel}}{{Charbonnel}}{1995}]{CHARBONNEL95}
{Charbonnel} C.,  1995, \mn@doi [\apjl] {10.1086/309744}, \href
  {https://ui.adsabs.harvard.edu/abs/1995ApJ...453L..41C} {453, L41}

\bibitem[\protect\citeauthoryear{{Charbonnel} \& {Lagarde}}{{Charbonnel} \&
  {Lagarde}}{2010}]{CL10}
{Charbonnel} C.,  {Lagarde} N.,  2010, \mn@doi [\aap]
  {10.1051/0004-6361/201014432}, \href
  {https://ui.adsabs.harvard.edu/abs/2010A&A...522A..10C} {522, A10}

\bibitem[\protect\citeauthoryear{{Charbonnel} \& {Zahn}}{{Charbonnel} \&
  {Zahn}}{2007}]{CZ07}
{Charbonnel} C.,  {Zahn} J.~P.,  2007, \mn@doi [\aap]
  {10.1051/0004-6361:20077274}, \href
  {https://ui.adsabs.harvard.edu/abs/2007A&A...467L..15C} {467, L15}

\bibitem[\protect\citeauthoryear{{Charbonnel} et~al.,}{{Charbonnel}
  et~al.}{2020}]{CHARBONNEL20}
{Charbonnel} C.,  et~al., 2020, \mn@doi [\aap] {10.1051/0004-6361/201936360},
  \href {https://ui.adsabs.harvard.edu/abs/2020A&A...633A..34C} {633, A34}

\bibitem[\protect\citeauthoryear{{Choi}, {Dotter}, {Conroy}, {Cantiello},
  {Paxton}  \& {Johnson}}{{Choi} et~al.}{2016}]{CHOI16}
{Choi} J.,  {Dotter} A.,  {Conroy} C.,  {Cantiello} M.,  {Paxton} B.,
  {Johnson} B.~D.,  2016, \mn@doi [\apj] {10.3847/0004-637X/823/2/102}, \href
  {https://ui.adsabs.harvard.edu/abs/2016ApJ...823..102C} {823, 102}

\bibitem[\protect\citeauthoryear{{Denissenkov}}{{Denissenkov}}{2010}]{DENISSENKOV10}
{Denissenkov} P.~A.,  2010, \mn@doi [\apj] {10.1088/0004-637X/723/1/563}, \href
  {https://ui.adsabs.harvard.edu/abs/2010ApJ...723..563D} {723, 563}

\bibitem[\protect\citeauthoryear{{Denissenkov} \& {Merryfield}}{{Denissenkov}
  \& {Merryfield}}{2011}]{DENISSENKOV11}
{Denissenkov} P.~A.,  {Merryfield} W.~J.,  2011, \mn@doi [\apjl]
  {10.1088/2041-8205/727/1/L8}, \href
  {https://ui.adsabs.harvard.edu/abs/2011ApJ...727L...8D} {727, L8}

\bibitem[\protect\citeauthoryear{{Denissenkov}, {Pinsonneault}  \&
  {MacGregor}}{{Denissenkov} et~al.}{2009}]{DENISSENKOV09}
{Denissenkov} P.~A.,  {Pinsonneault} M.,   {MacGregor} K.~B.,  2009, \mn@doi
  [\apj] {10.1088/0004-637X/696/2/1823}, \href
  {https://ui.adsabs.harvard.edu/abs/2009ApJ...696.1823D} {696, 1823}

\bibitem[\protect\citeauthoryear{{Dias}, {Alessi}, {Moitinho}  \&
  {L{\'e}pine}}{{Dias} et~al.}{2002}]{DIAS02}
{Dias} W.~S.,  {Alessi} B.~S.,  {Moitinho} A.,   {L{\'e}pine} J.~R.~D.,  2002,
  \mn@doi [\aap] {10.1051/0004-6361:20020668}, \href
  {https://ui.adsabs.harvard.edu/abs/2002A&A...389..871D} {389, 871}

\bibitem[\protect\citeauthoryear{{Donor} et~al.,}{{Donor}
  et~al.}{2018}]{DONOR18}
{Donor} J.,  et~al., 2018, \mn@doi [\aj] {10.3847/1538-3881/aad635}, \href
  {https://ui.adsabs.harvard.edu/abs/2018AJ....156..142D} {156, 142}

\bibitem[\protect\citeauthoryear{{Donor} et~al.,}{{Donor}
  et~al.}{2020}]{DONOR20}
{Donor} J.,  et~al., 2020, \mn@doi [\aj] {10.3847/1538-3881/ab77bc}, \href
  {https://ui.adsabs.harvard.edu/abs/2020AJ....159..199D} {159, 199}

\bibitem[\protect\citeauthoryear{{Dotter}}{{Dotter}}{2016}]{DOTTER16}
{Dotter} A.,  2016, \mn@doi [\apjs] {10.3847/0067-0049/222/1/8}, \href
  {https://ui.adsabs.harvard.edu/abs/2016ApJS..222....8D} {222, 8}

\bibitem[\protect\citeauthoryear{{Drazdauskas}, {Tautvai{\v{s}}ien{\.{e}}},
  {Randich}, {Bragaglia}, {Mikolaitis}  \& {Janulis}}{{Drazdauskas}
  et~al.}{2016}]{DRAZDAUSKAS16}
{Drazdauskas} A.,  {Tautvai{\v{s}}ien{\.{e}}} G.,  {Randich} S.,  {Bragaglia}
  A.,  {Mikolaitis} {\v{S}}.,   {Janulis} R.,  2016, \mn@doi [\aap]
  {10.1051/0004-6361/201628138}, \href
  {https://ui.adsabs.harvard.edu/abs/2016A&A...589A..50D} {589, A50}

\bibitem[\protect\citeauthoryear{{Frinchaboy} et~al.,}{{Frinchaboy}
  et~al.}{2013}]{FRINCHABOY13}
{Frinchaboy} P.~M.,  et~al., 2013, \mn@doi [\apjl]
  {10.1088/2041-8205/777/1/L1}, \href
  {https://ui.adsabs.harvard.edu/abs/2013ApJ...777L...1F} {777, L1}

\bibitem[\protect\citeauthoryear{{Garc{\'{\i}}a P{\'e}rez}
  et~al.,}{{Garc{\'{\i}}a P{\'e}rez} et~al.}{2016}]{GarciaPerez2016}
{Garc{\'{\i}}a P{\'e}rez} A.~E.,  et~al., 2016, \mn@doi [\aj]
  {10.3847/0004-6256/151/6/144}, \href
  {http://adsabs.harvard.edu/abs/2016AJ....151..144G} {151, 144}

\bibitem[\protect\citeauthoryear{{Gilroy}}{{Gilroy}}{1989}]{GILROY89}
{Gilroy} K.~K.,  1989, \mn@doi [\apj] {10.1086/168173}, \href
  {https://ui.adsabs.harvard.edu/abs/1989ApJ...347..835G} {347, 835}

\bibitem[\protect\citeauthoryear{{Gratton}, {Sneden}, {Carretta}  \&
  {Bragaglia}}{{Gratton} et~al.}{2000}]{GRATTON00}
{Gratton} R.~G.,  {Sneden} C.,  {Carretta} E.,   {Bragaglia} A.,  2000, \aap,
  \href {https://ui.adsabs.harvard.edu/abs/2000A&A...354..169G} {354, 169}

\bibitem[\protect\citeauthoryear{{Gunn} et~al.,}{{Gunn}
  et~al.}{2006}]{Gunn2006}
{Gunn} J.~E.,  et~al., 2006, \mn@doi [\aj] {10.1086/500975}, \href
  {http://adsabs.harvard.edu/abs/2006AJ....131.2332G} {131, 2332}

\bibitem[\protect\citeauthoryear{{Gustafsson}, {Edvardsson}, {Eriksson},
  {J{\o}rgensen}, {Nordlund}  \& {Plez}}{{Gustafsson}
  et~al.}{2008}]{GUSTAFSSON08}
{Gustafsson} B.,  {Edvardsson} B.,  {Eriksson} K.,  {J{\o}rgensen} U.~G.,
  {Nordlund} {\r{A}}.,   {Plez} B.,  2008, \mn@doi [\aap]
  {10.1051/0004-6361:200809724}, \href
  {https://ui.adsabs.harvard.edu/abs/2008A&A...486..951G} {486, 951}

\bibitem[\protect\citeauthoryear{{Hayes} et~al.,}{{Hayes} et~al.}{2022}]{HAYES}
{Hayes} C.~R.,  et~al., 2022, arXiv e-prints, \href
  {https://ui.adsabs.harvard.edu/abs/2022arXiv220800071H} {p. arXiv:2208.00071}

\bibitem[\protect\citeauthoryear{{Holtzman}, {Harrison}  \&
  {Coughlin}}{{Holtzman} et~al.}{2010}]{Holtzman2010}
{Holtzman} J.~A.,  {Harrison} T.~E.,   {Coughlin} J.~L.,  2010, \mn@doi
  [Advances in Astronomy] {10.1155/2010/193086}, \href
  {http://adsabs.harvard.edu/abs/2010AdAst2010E..46H} {2010, 193086}

\bibitem[\protect\citeauthoryear{{Holtzman} et~al.,}{{Holtzman}
  et~al.}{2015}]{Holtzman2015}
{Holtzman} J.~A.,  et~al., 2015, \mn@doi [\aj] {10.1088/0004-6256/150/5/148},
  \href {http://adsabs.harvard.edu/abs/2015AJ....150..148H} {150, 148}

\bibitem[\protect\citeauthoryear{{J{\"o}nsson} et~al.,}{{J{\"o}nsson}
  et~al.}{2020}]{JONSSON20}
{J{\"o}nsson} H.,  et~al., 2020, \mn@doi [\aj] {10.3847/1538-3881/aba592},
  \href {https://ui.adsabs.harvard.edu/abs/2020AJ....160..120J} {160, 120}

\bibitem[\protect\citeauthoryear{{Kippenhahn}, {Ruschenplatt}  \&
  {Thomas}}{{Kippenhahn} et~al.}{1980}]{KIPPENHAHN80}
{Kippenhahn} R.,  {Ruschenplatt} G.,   {Thomas} H.~C.,  1980, \aap, \href
  {https://ui.adsabs.harvard.edu/abs/1980A&A....91..175K} {91, 175}

\bibitem[\protect\citeauthoryear{{Lagarde}, {Decressin}, {Charbonnel},
  {Eggenberger}, {Ekstr{\"o}m}  \& {Palacios}}{{Lagarde}
  et~al.}{2012}]{LAGARDE12}
{Lagarde} N.,  {Decressin} T.,  {Charbonnel} C.,  {Eggenberger} P.,
  {Ekstr{\"o}m} S.,   {Palacios} A.,  2012, \mn@doi [\aap]
  {10.1051/0004-6361/201118331}, \href
  {https://ui.adsabs.harvard.edu/abs/2012A&A...543A.108L} {543, A108}

\bibitem[\protect\citeauthoryear{{Lagarde} et~al.,}{{Lagarde}
  et~al.}{2019}]{LAGARDE19}
{Lagarde} N.,  et~al., 2019, \mn@doi [\aap] {10.1051/0004-6361/201732433},
  \href {https://ui.adsabs.harvard.edu/abs/2019A&A...621A..24L} {621, A24}

\bibitem[\protect\citeauthoryear{{Maeder}, {Meynet}, {Lagarde}  \&
  {Charbonnel}}{{Maeder} et~al.}{2013}]{MAEDER13}
{Maeder} A.,  {Meynet} G.,  {Lagarde} N.,   {Charbonnel} C.,  2013, \mn@doi
  [\aap] {10.1051/0004-6361/201220936}, \href
  {https://ui.adsabs.harvard.edu/abs/2013A&A...553A...1M} {553, A1}

\bibitem[\protect\citeauthoryear{{Majewski} et~al.,}{{Majewski}
  et~al.}{2017}]{MAJEWSKI17}
{Majewski} S.~R.,  et~al., 2017, \mn@doi [\aj] {10.3847/1538-3881/aa784d},
  \href {https://ui.adsabs.harvard.edu/abs/2017AJ....154...94M} {154, 94}

\bibitem[\protect\citeauthoryear{{Masseron}, {Merle}  \& {Hawkins}}{{Masseron}
  et~al.}{2016}]{BACCHUS}
{Masseron} T.,  {Merle} T.,   {Hawkins} K.,  2016, {BACCHUS: Brussels Automatic
  Code for Characterizing High accUracy Spectra}, Astrophysics Source Code
  Library, record ascl:1605.004 (\mn@eprint {ascl} {1605.004})

\bibitem[\protect\citeauthoryear{{Myers} et~al.,}{{Myers} et~al.}{2022}]{MYERS}
{Myers} N.,  et~al., 2022, \mn@doi [\aj] {10.3847/1538-3881/ac7ce5}, \href
  {https://ui.adsabs.harvard.edu/abs/2022AJ....164...85M} {164, 85}

\bibitem[\protect\citeauthoryear{{Nidever} et~al.,}{{Nidever}
  et~al.}{2015}]{dln15}
{Nidever} D.~L.,  et~al., 2015, \mn@doi [\aj] {10.1088/0004-6256/150/6/173},
  \href {http://adsabs.harvard.edu/abs/2015AJ....150..173N} {150, 173}

\bibitem[\protect\citeauthoryear{{Palacios}, {Talon}, {Charbonnel}  \&
  {Forestini}}{{Palacios} et~al.}{2003}]{PALACIOS03}
{Palacios} A.,  {Talon} S.,  {Charbonnel} C.,   {Forestini} M.,  2003, \mn@doi
  [\aap] {10.1051/0004-6361:20021759}, \href
  {https://ui.adsabs.harvard.edu/abs/2003A&A...399..603P} {399, 603}

\bibitem[\protect\citeauthoryear{{Palacios}, {Charbonnel}, {Talon}  \&
  {Siess}}{{Palacios} et~al.}{2006}]{PALACIOS06}
{Palacios} A.,  {Charbonnel} C.,  {Talon} S.,   {Siess} L.,  2006, \mn@doi
  [\aap] {10.1051/0004-6361:20053065}, \href
  {https://ui.adsabs.harvard.edu/abs/2006A&A...453..261P} {453, 261}

\bibitem[\protect\citeauthoryear{{Paxton}, {Bildsten}, {Dotter}, {Herwig},
  {Lesaffre}  \& {Timmes}}{{Paxton} et~al.}{2011}]{PAXTON11}
{Paxton} B.,  {Bildsten} L.,  {Dotter} A.,  {Herwig} F.,  {Lesaffre} P.,
  {Timmes} F.,  2011, \mn@doi [\apjs] {10.1088/0067-0049/192/1/3}, \href
  {https://ui.adsabs.harvard.edu/abs/2011ApJS..192....3P} {192, 3}

\bibitem[\protect\citeauthoryear{{Paxton} et~al.,}{{Paxton}
  et~al.}{2013}]{PAXTON13}
{Paxton} B.,  et~al., 2013, \mn@doi [\apjs] {10.1088/0067-0049/208/1/4}, \href
  {https://ui.adsabs.harvard.edu/abs/2013ApJS..208....4P} {208, 4}

\bibitem[\protect\citeauthoryear{{Paxton} et~al.,}{{Paxton}
  et~al.}{2015}]{PAXTON15}
{Paxton} B.,  et~al., 2015, \mn@doi [\apjs] {10.1088/0067-0049/220/1/15}, \href
  {https://ui.adsabs.harvard.edu/abs/2015ApJS..220...15P} {220, 15}

\bibitem[\protect\citeauthoryear{{Paxton} et~al.,}{{Paxton}
  et~al.}{2018}]{PAXTON18}
{Paxton} B.,  et~al., 2018, \mn@doi [\apjs] {10.3847/1538-4365/aaa5a8}, \href
  {https://ui.adsabs.harvard.edu/abs/2018ApJS..234...34P} {234, 34}

\bibitem[\protect\citeauthoryear{{Sackmann} \& {Boothroyd}}{{Sackmann} \&
  {Boothroyd}}{1999}]{SB99}
{Sackmann} I.~J.,  {Boothroyd} A.~I.,  1999, \mn@doi [\apj] {10.1086/306545},
  \href {https://ui.adsabs.harvard.edu/abs/1999ApJ...510..217S} {510, 217}

\bibitem[\protect\citeauthoryear{{Santana} et~al.,}{{Santana}
  et~al.}{2021}]{Santana2021}
{Santana} F.~A.,  et~al., 2021, \mn@doi [\aj] {10.3847/1538-3881/ac2cbc}, \href
  {https://ui.adsabs.harvard.edu/abs/2021AJ....162..303S} {162, 303}

\bibitem[\protect\citeauthoryear{{Sengupta} \& {Garaud}}{{Sengupta} \&
  {Garaud}}{2018}]{SG18}
{Sengupta} S.,  {Garaud} P.,  2018, \mn@doi [\apj] {10.3847/1538-4357/aacbc8},
  \href {https://ui.adsabs.harvard.edu/abs/2018ApJ...862..136S} {862, 136}

\bibitem[\protect\citeauthoryear{{Shetrone} et~al.,}{{Shetrone}
  et~al.}{2015}]{shetrone2015}
{Shetrone} M.,  et~al., 2015, \mn@doi [\apjs] {10.1088/0067-0049/221/2/24},
  \href {http://adsabs.harvard.edu/abs/2015ApJS..221...24S} {221, 24}

\bibitem[\protect\citeauthoryear{{Smiljanic}, {Gauderon}, {North}, {Barbuy},
  {Charbonnel}  \& {Mowlavi}}{{Smiljanic} et~al.}{2009}]{SMILJANIC09}
{Smiljanic} R.,  {Gauderon} R.,  {North} P.,  {Barbuy} B.,  {Charbonnel} C.,
  {Mowlavi} N.,  2009, \mn@doi [\aap] {10.1051/0004-6361/200811113}, \href
  {https://ui.adsabs.harvard.edu/abs/2009A&A...502..267S} {502, 267}

\bibitem[\protect\citeauthoryear{{Smith} et~al.,}{{Smith}
  et~al.}{2021}]{Smith2021}
{Smith} V.~V.,  et~al., 2021, \mn@doi [\aj] {10.3847/1538-3881/abefdc}, \href
  {https://ui.adsabs.harvard.edu/abs/2021AJ....161..254S} {161, 254}

\bibitem[\protect\citeauthoryear{{Sneden}, {Pilachowski}  \&
  {Vandenberg}}{{Sneden} et~al.}{1986}]{SNEDEN86}
{Sneden} C.,  {Pilachowski} C.~A.,   {Vandenberg} D.~A.,  1986, \mn@doi [\apj]
  {10.1086/164822}, \href
  {https://ui.adsabs.harvard.edu/abs/1986ApJ...311..826S} {311, 826}

\bibitem[\protect\citeauthoryear{{Stern}}{{Stern}}{1960}]{STERN60}
{Stern} M.~E.,  1960, \mn@doi [Tellus] {10.3402/tellusa.v12i2.9378}, \href
  {https://ui.adsabs.harvard.edu/abs/1960Tell...12..172S} {12, 172}

\bibitem[\protect\citeauthoryear{{Sweigart} \& {Mengel}}{{Sweigart} \&
  {Mengel}}{1979}]{SWEIGART79}
{Sweigart} A.~V.,  {Mengel} J.~G.,  1979, \mn@doi [\apj] {10.1086/156996},
  \href {https://ui.adsabs.harvard.edu/abs/1979ApJ...229..624S} {229, 624}

\bibitem[\protect\citeauthoryear{{Szigeti} et~al.,}{{Szigeti}
  et~al.}{2018}]{SZIGETI18}
{Szigeti} L.,  et~al., 2018, \mn@doi [\mnras] {10.1093/mnras/stx3027}, \href
  {https://ui.adsabs.harvard.edu/abs/2018MNRAS.474.4810S} {474, 4810}

\bibitem[\protect\citeauthoryear{{Takeda}, {Omiya}, {Harakawa}  \&
  {Sato}}{{Takeda} et~al.}{2019}]{TAKEDA19}
{Takeda} Y.,  {Omiya} M.,  {Harakawa} H.,   {Sato} B.,  2019, \mn@doi [\pasj]
  {10.1093/pasj/psz104}, \href
  {https://ui.adsabs.harvard.edu/abs/2019PASJ...71..119T} {71, 119}

\bibitem[\protect\citeauthoryear{{Tautvai{\v{s}}ien{\.{e}}}, {Edvardsson},
  {Puzeras}, {Barisevi{\v{c}}ius}  \& {Ilyin}}{{Tautvai{\v{s}}ien{\.{e}}}
  et~al.}{2010}]{TAUTVAISIENE10}
{Tautvai{\v{s}}ien{\.{e}}} G.,  {Edvardsson} B.,  {Puzeras} E.,
  {Barisevi{\v{c}}ius} G.,   {Ilyin} I.,  2010, \mn@doi [\mnras]
  {10.1111/j.1365-2966.2010.17381.x}, \href
  {https://ui.adsabs.harvard.edu/abs/2010MNRAS.409.1213T} {409, 1213}

\bibitem[\protect\citeauthoryear{{Tautvai{\v{s}}ien{\.{e}}},
  {Barisevi{\v{c}}ius}, {Chorniy}, {Ilyin}  \&
  {Puzeras}}{{Tautvai{\v{s}}ien{\.{e}}} et~al.}{2013}]{TAUTVAISIENE13}
{Tautvai{\v{s}}ien{\.{e}}} G.,  {Barisevi{\v{c}}ius} G.,  {Chorniy} Y.,
  {Ilyin} I.,   {Puzeras} E.,  2013, \mn@doi [\mnras] {10.1093/mnras/sts663},
  \href {https://ui.adsabs.harvard.edu/abs/2013MNRAS.430..621T} {430, 621}

\bibitem[\protect\citeauthoryear{{Tayar} \& {Joyce}}{{Tayar} \&
  {Joyce}}{2022}]{TAYAR22}
{Tayar} J.,  {Joyce} M.,  2022, \mn@doi [\apjl] {10.3847/2041-8213/ac85ab},
  \href {https://ui.adsabs.harvard.edu/abs/2022ApJ...935L..30T} {935, L30}

\bibitem[\protect\citeauthoryear{{Traxler}, {Garaud}  \& {Stellmach}}{{Traxler}
  et~al.}{2011}]{TRAXLER11}
{Traxler} A.,  {Garaud} P.,   {Stellmach} S.,  2011, \mn@doi [\apjl]
  {10.1088/2041-8205/728/2/L29}, \href
  {https://ui.adsabs.harvard.edu/abs/2011ApJ...728L..29T} {728, L29}

\bibitem[\protect\citeauthoryear{{Ulrich}}{{Ulrich}}{1972}]{ULRICH72}
{Ulrich} R.~K.,  1972, \mn@doi [\apj] {10.1086/151336}, \href
  {https://ui.adsabs.harvard.edu/abs/1972ApJ...172..165U} {172, 165}

\bibitem[\protect\citeauthoryear{{Wasserburg}, {Boothroyd}  \&
  {Sackmann}}{{Wasserburg} et~al.}{1995}]{WBS95}
{Wasserburg} G.~J.,  {Boothroyd} A.~I.,   {Sackmann} I.~J.,  1995, \mn@doi
  [\apjl] {10.1086/309555}, \href
  {https://ui.adsabs.harvard.edu/abs/1995ApJ...447L..37W} {447, L37}

\bibitem[\protect\citeauthoryear{{Wilson} et~al.,}{{Wilson}
  et~al.}{2019}]{wilson2019}
{Wilson} J.~C.,  et~al., 2019, \mn@doi [\pasp] {10.1088/1538-3873/ab0075},
  \href {https://ui.adsabs.harvard.edu/abs/2019PASP..131e5001W} {131, 055001}

\bibitem[\protect\citeauthoryear{{Zasowski} et~al.,}{{Zasowski}
  et~al.}{2013}]{zas13}
{Zasowski} G.,  et~al., 2013, \mn@doi [\aj] {10.1088/0004-6256/146/4/81}, \href
  {https://ui.adsabs.harvard.edu/abs/2013AJ....146...81Z} {146, 81}

\bibitem[\protect\citeauthoryear{{Zasowski} et~al.,}{{Zasowski}
  et~al.}{2017}]{zas17}
{Zasowski} G.,  et~al., 2017, \mn@doi [\aj] {10.3847/1538-3881/aa8df9}, \href
  {http://adsabs.harvard.edu/abs/2017AJ....154..198Z} {154, 198}

\makeatother
\end{thebibliography}

%%%%%%%%%%%%%%%%%%%%%%%%%%%%%%%%%%%%%%%%%%%%%%%%%%

% Don't change these lines
\bsp	% typesetting comment
\label{lastpage}
\end{document}